\documentclass[useAMS,usenatbib,usegraphicx]{mn2e}
\usepackage{verbatim}
\newcommand{\bugref}{\bibitem[\protect\citeauthoryear{dummy }{1893}]{dum}}

\title[Faraday Rotation Gradients in AGN]
{Connecting Magnetic Towers with Faraday Rotation Gradients in 
Active Galactic Nuclei Jets.}

\author[M. Mahmud, C. P. Coughlan, E. Murphy, D. C. Gabuzda \& D. R. Hallahan]
{M. Mahmud$^1$, C. P. Coughlan$^2$, E. Murphy$^2$, D. C. Gabuzda$^2$ \& D. R. Hallahan$^2$ \\
$^1$Joint Institute for VLBI in Europe, Postbus 2, Dwingeloo 7900 AA, The Netherlands\\
$^2$Physics Department, University College Cork, Cork, Ireland} 

\begin{document}
\date{}
\pagerange{\pageref{firstpage}--\pageref{lastpage}} \pubyear{2013}
\maketitle
\label{firstpage}

\begin{abstract}

The idea that systematic Faraday Rotation gradients across the parsec-scale 
jets of Active Galactic Nuclei (AGNs) can reveal the presence of helical 
magnetic ({\bf B}) fields has been around since the early 1990s, although
the first observation of this phenomenon was about ten years later. These 
gradients are taken to be due to the systematic variation of the line-of-sight 
{\bf B} field across the jet. We present here the parsec-scale Faraday Rotation 
distributions for the BL~Lac objects 0716+714 
and 1749+701, based on polarization data obtained with 
the Very Long Baseline Array (VLBA) at two wavelengths near each of the 
2cm, 4cm and 6cm bands (0716+714) and at four wavelengths in the range 
18--22 cm (1749+701). The 
Rotation Measure (RM) maps for both these sources indicate systematic gradients 
across their jets, as expected if these jets have helical {\bf B} fields.
The significance of these transverse RM gradients is 
$> 3\sigma$ in all cases.  We present the results of Monte Carlo 
simulations directly demonstrating the possibility of observing such
transverse RM gradients even if the intrinsic jet structure is much
narrower than the observing beam.
We observe an intriguing new feature in these sources, a reversal 
in the direction of the gradient in the jet as compared to the gradient in 
the core region.  This provides new evidence to support models in which 
field lines emerging from the central region of the accretion disk and
closing in the outer region of the accretion disk are both ``wound up'' by 
the differential rotation of the disk. The net observed RM gradient will 
essentially be the sum effect of two regions of helical field, one nested 
inside the other. The direction of the net RM gradient will be determined by 
whether the inner or outer helix dominates the RM integrated through the jet, 
and RM gradient reversals will be observed if the inner and outer helical 
fields dominate in different regions of the jet. This potentially provides new 
insights about the overall configuration of the jet {\bf B} fields. 
\end{abstract}

\section{Introduction}

The radio emission of Active Galactic Nuclei (AGNs) is synchrotron radiation 
generated in the relativistic jets that emerge from the nucleus of the galaxy, 
presumably along the rotational axis of a central supermassive black hole. 
Synchrotron radiation can be highly linearly polarized, up to $\simeq 75\%$ in 
the case of a uniform magnetic ({\bf B}) field (Pacholczyk 1970). Linear 
polarization observations are essential, as they give information about the 
orientation and degree of order of the {\bf B} field, as well as the 
distribution of thermal electrons and the {\bf B}-field geometry in the 
vicinity of the AGN. Many theorists have suggested that the magnetic fields of 
these sources are closely connected with the collimation of the jets, and 
could determine whether sources have prominent jets or not (eg. Meier et al. 
2001). Thus, information on the magnetic fields of these sources is essential 
in helping us better understand various physical processes in AGN jets.

VLBI polarization observations of BL~Lac objects have shown a tendency for 
the polarization {\bf E} vectors in the parsec-scale jets to be aligned with 
the local jet direction, which implies that the corresponding {\bf B} field 
is transverse to the jet, because the jet is optically thin (Gabuzda, 
Pushkarev \& Cawthorne 2000). It seems likely that many of these transverse 
{\bf B} fields represent the ordered toroidal component of the intrinsic 
{\bf B} fields of the jets, as discussed by Gabuzda et al. (2008), see also 
references therein. Depending on the observer's viewing angle and the helix's
pitch angle, helical jet 
{\bf B} fields can also give rise to a `spine-sheath' polarization structure 
in the frame of the observer, with a region of longitudinal polarization 
(transverse {\bf B}-vectors) along the central `spine' of the jet surrounded 
by regions of transverse polarization (longitudinal {\bf B}-vectors) near the 
edges of the jet. The presence of transverse polarization near the edges of 
the jet could  be a natural consequence of a helical jet {\bf B} field, 
although it has also been suggested to be due to interaction with the 
surrounding medium (Laing 1996; Lyutikov, Pariev \& Gabuzda 2005; Attridge, 
Roberts \& Wardle 1999; Pushkarev et al. 2005).

Faraday Rotation studies can play a key role in determining the intrinsic {\bf B} 
field geometries associated with the jets. Faraday Rotation of the plane of 
linear polarization occurs during the passage of an electromagnetic wave 
through a region with free electrons and a magnetic field with a non-zero 
component along the line-of-sight. The amount of rotation is proportional to 
the integral of the density of free electrons $n_{e}$ multiplied by the 
line-of-sight {\bf B} field, the square of the observing wavelength $\lambda^{2}$, and 
various physical constants; the coefficient of $\lambda^{2}$  is called the 
Rotation Measure (RM):
\begin{eqnarray}
           \Delta\chi\propto\lambda^{2}\int n_{e} B\cdot dl\equiv RM\lambda^{2}
\end{eqnarray}
The intrinsic polarization angle can be obtained from the relation:
\begin{eqnarray}
           \chi_{obs} =  \chi_0 + RM \lambda^{2}
\end{eqnarray}
where $\chi_{obs}$ is the observed polarization angle, $\chi_0$ is the 
intrinsic polarization angle in the absence of Faraday rotation and $\lambda$ 
is the observing wavelength. Simultaneous multifrequency 
observations thus potentially enable the determination of the RM, as well as identification 
of the intrinsic polarization angles.

Systematic gradients in the Faraday Rotation Measure (RM) have been reported 
previously across the parsec-scale jets of several AGNs, interpreted as 
reflecting the systematic change in the line-of-sight component of a toroidal 
or helical jet {\bf B} field across the jet (Blandford 1993, Asada et al. 2002, 
Gabuzda, Murray \& Cronin 2004, Zavala \& Taylor 2005, Gabuzda et al. 2008, 
Asada et al. 2008a,b,2010, Mahmud, Gabuzda \& Bezrukovs 2009). Such fields would 
come about in a natural way as a result of the `winding up' of an initial 
`seed' field by the differential rotation of the central accreting objects 
(e.g. Nakamura et al.  2001, Lovelace et al. 2002). 

We consider here two objects in which we have detected transverse RM 
gradients in both the core region and jet: 0716+714 and 1749+701. In both cases, 
there is a reversal of the direction of the RM gradient between these two
regions. We discuss a 
possible explanation of this phenomena based on 
magnetic-tower-type models for jet launching. Throughout we assume $H_{0}$ = 
71 km/s/Mpc, {$\Omega_{\lambda}$ = 0.73} and {$\Omega_{m}$ = 0.27}.

\section{Faraday-Rotation Observations and Reduction}
Very Long Baseline Array (VLBA) polarization observations of the sources 
included in this paper were carried out as part of two different studies of 
the same sample of BL Lac objects: one at 4.6--15.4~GHz and one at 1.36--1.67~GHz. 
The high-frequency observations of 0716+714 were on 22 March 2004 and of
1749+701 were on 22 August 2003; the low-frequency observations of 1749+701
were on 17 January 2004. In both cases, the distributions of $u$--$v$ points
were virtually identical for the different frequencies observed during a single
set of observations, with the baseline lengths scaled in accordance with the
individual observing frequencies.

Standard tasks in the NRAO AIPS package were used for the amplitude calibration 
and preliminary phase calibration. The instrumental polarizations (`D-Terms') 
were determined with the task `LPCAL', solving simultaneously for the 
source polarization.  In all cases, the reference antenna used was Los Alamos. 

The Electric Vector Position Angle (EVPA) calibration was done using 
integrated polarization observations of bright, compact sources, obtained with 
the Very Large Array (VLA) near in time to our VLBA observations, by rotating 
the EVPA for the total VLBI polarization of the source to match the EVPA for 
the integrated polarization of that source derived from VLA observations. 

\subsection{August 2003 and March 2004 Observations: 4.6--15.4~GHz}
The observations were carried out at six frequencies: 4.612, 5.092, 7.916, 
8.883, 12.939 and 15.383~GHz. Each source was observed for about 25--30 minutes 
at each frequency, in a `snap-shot' mode with 8--10 scans spread out over the 
observing time period. Presented in this paper are the results for 
0716+714 (observed on 22 March 2004) and 1749+701 (observed on 22 August 2003).

The instrumental polarizations (`D-Terms') were determined using observations 
of 1156+295 (22 August 2003) and 0235+164 (22 March 2004). 
The source of the integrated VLA polarizations for the EVPA calibration
was the NRAO website (www.aoc.nrao.edu/~smyers/calibration/).  The VLA 
observations were made at frequencies 5, 8.5, 22 and 43 GHz. We found these
EVPA values to be consistent with a linear $\lambda^{2}$ law (Faraday Rotation) 
and were thus able to interpolate the corresponding values for our 
non-standard frequencies (see Mahmud et al. 2009).  The sources used were 
1803+784 and 2200+420.  We refined 
our initial EVPA calibration by examining the resulting polarization images 
for several sources with simple structures and checking for consistency. This 
led to adjustments of $5-20^{\circ}$ for several of the EVPA corrections. This 
procedure improved the overall self consistency of the polarization and RM 
maps for virtually all of the sources observed. 
%Due to the stability of the 
%required EVPA correction for a given reference antenna over periods of several 
%years (e.g. Reynolds et al. 2001), we were able to apply this refined EVPA 
%calibration to all the epochs of our observations. 
We estimate that our 
overall EVPA calibration is accurate to within $3^{\circ}$. 
A summary of our final 4.6--15.4 GHz EVPA 
corrections is given in Mahmud et al. (2009). 

\begin{table}
\caption{EVPA calibrations for 17 January 2004} \centering
\label{tab:EVPA_red}
\begin{tabular}{ll}
\hline
Frequency (GHz)           & $\Delta\chi$ (deg)                \\\hline
1.358                     & 128.1   \\
1.430                     & 111.1   \\
1.493                     & 100.4   \\
1.665                     & 82.5   \\
\hline
\end{tabular}
\end{table}
\hfill

To verify the accuracy of the overall flux calibration at 4.6--15.4 GHz, we 
determined the spectra of various optically thin regions in the jet of 
1803+784, after taking into account the relative shifts between the images
(see Mahmud et al. 2009). The observed 
4.6--15.4~GHz fluxes are consistent with a power-law within the errors, 
corresponding to a ``normal'' optically thin spectral indices of $\simeq$1. 

\subsection{January 2004 Observations: 1.36--1.67~GHz}
Results for 1749+701 at epoch 17 January 2004 at four frequencies between 
1.358 and 1.665 GHz are also included in this paper. The instrumental 
polarizations (`D-Terms') for these  observations were determined using 
observations of 0851+202. The absolute calibration of the EVPAs was  
determined using VLA observations of 0851+202 obtained at 1.485 and 1.665~GHz 
on February 20, 2004. These observations were sufficiently close in time 
to the VLBA observations to be suitable for the EVPA calibration because 
the polarization is not rapidly variable at such low frequencies. A 
lambda-squared fit was applied to the VLA polarization angles, yielding a 
rotation measure of +31.6~rad/m$^2$, in excellent agreement with the previously 
measured value of $+31 \pm 2$ (Pushkarev 2001). We accordingly used the 
measured VLA polarization angles and rotation measure to determine the 
integrated polarization angles for our four VLBA frequencies, which were then 
used to calibrate the  VLBA EVPAs. We estimate the errors in the resulting 
polarization angles to be no more than 2$^\circ$. For a list of the observing 
frequencies and their EVPA corrections, see Table~\ref{tab:EVPA_red}.

\subsection{Rotation Measure Determination}
We made maps of the distribution of the total intensity $I$ and Stokes 
parameters $Q$ and $U$ at each of the frequencies, with matched cell sizes,
images sizes, and resolutions, by convolving all of the final maps with 
the same beam.  The distributions of 
the polarized flux ($p = \sqrt{Q^2 + U^2}$), as well as maps of the EVPA 
($\chi = \frac{1}{2}\arctan \frac{U}{Q}$) and $\chi$ noise maps, were obtained 
from the $Q$ and $U$ maps using the task `COMB'. 

Although the $Q$ and $U$ maps at each frequency will be properly aligned 
with the $I$ map at that same frequency, the images at different frequencies 
can be appreciably shifted relative to each other. The physical origin of this 
effect is the fact that the mapping procedure effectively aligns the images 
roughly on the bright, compact VLBI core, whose position depends on 
the observing frequency: the core ($\tau = 1$ surface) appears further down 
the jet at lower frequencies (K\"onigl 1981). It is important to correct for 
this effect by properly aligning the $I$ images before making spectral-index 
maps; although the effect of modest misalignments between the $\chi$ maps 
at different frequencies is much smaller, it is nevertheless optimal to align
these maps before constructing RM maps to maximize the reliability of the 
resulting RM maps. 

We determined the relative shifts between the maps at each of our frequencies using
the cross-correlation algorithm of Croke \& Gabuzda (2008), which essentially
aligns the images based on their optically thin jet emission. This procedure yielded
only negligible ``core-shifts'' for 0716+714 in our frequency range (less than a 
pixel), consistent with the results of Kovalev et al. (2008). The core-shift for
1749+701 between 4.6~GHz and 15.4~GHz is appreciable: approximately 0.8~mas in 
position angle $-63^{\circ}$, aligned with the VLBI jet direction. The
shifts between the other frequencies and 15.4~GHz were smaller and consistent with
the expected scaling with frequency. We accordingly used these shifts to 
align the polarization-angle maps for 1749+701 before making the high-frequency
RM map. We verified that the shifts between the images at the lower frequencies
(1.36--1.67~GHz) were neglible for all sources observed in that experiment, 
including 1749+701; this is expected, since these frequencies do not cover
a wide range; accordingly, no shifts were necessary for those images. 

Further, we constructed maps of the RM, using the AIPS task `RM',
after first subtracting the effect of the integrated RM (Pushkarev 2001), presumed 
to arise in our Galaxy, from the observed polarization 
angles, so that any residual Faraday Rotation was due to only 
the thermal plasma in the vicinity of the AGN. We used a modified version
of `RM' enabling simultaneously RM fitting using 
up to eight frequencies.  We used the option in the task 
`RM' of blanking output pixels when the uncertainty in the RM exceeds a 
specified value, which was about 30~rad/m$^{2}$ for the high-frequency maps
and about 10~rad/m$^2$ for the low-frequency maps. This uncertainty in the
RM calculated for a given pixel by the `RM' task is based on a fit of $\chi$ vs.
$\lambda^2$ weighted by the uncertainties in the polarization angles, which are,
in turn, calculated using the noise levels on the Stokes $Q$ and $U$ maps. Thus,
the resulting RM uncertainties for individual pixels are determined both by the
uncertainties in the polarization angles and the quality of the linear $\lambda^2$
fit.  The blanking levels we chose
were determined empirically, as the maximum values that did not lead to any
obviously spurious features in the RM maps, either at the location of the 
source or in the rest of the map.  The blanking applied essentially means that
the output RM values show some evidence of reality and reliability -- 
colocation with the source emission region and agreement with a $\lambda^2$
law within the specified limit. 

\section{Estimation of the $\chi$ and RM Uncertainties}

It has been usual to adopt the root-mean-square (rms) deviations in the residual 
map (or in the final CLEAN map far from any regions containing real flux)
$\sigma_{rms}$ as an estimate of the total uncertainty in the measured flux in
an individual pixel. Hovatta et al. (2012) have recently investigated this
practice empirically using Monte Carlo simulations.  They concluded that
the uncertainties in $Q$ and $U$ fluxes in individual pixels are described
well by the expression

\begin{equation}
\sigma = \sqrt{\sigma_{rms}^2 + \sigma_{Dterm}^2 + (1.5\sigma_{rms})^2}
\end{equation}

\noindent
where $\sigma_{Dterm}$ is 
associated with the presence of residual instrumental polarizations in the data
(see also Roberts, Wardle \& Brown 1994):

\begin{equation}
\sigma_{Dterm} 
\simeq
\frac{\sigma_{\Delta}}{\sqrt{N_{ant} N_{IF} N_{scan}}}\sqrt{I^2 + (0.3\, I_{peak})^2}
\end{equation}

\noindent
where $\sigma_{\Delta}$ is the estimated uncertainty in the individual D-terms, 
$N_{ant}$ the number of antennas in the VLB array (assuming all have 
altitude-azimuth mounts), $N_{IF}$ the number of IFs (sub-bands within the 
total observed band at a given frequency) used for the observations, 
$N_{scan}$ the number of scans with independent parallactic angles, $I$ the 
total intensity at the point in question, and $I_{peak}$ the total intensity 
at the map peak. The term containing $I_{peak}$ was added by Hovatta 
et al. (2012) to approximately take into account the fact that the residual 
D-term uncertainty tends to scatter polarized flux throughout the map. 
The expression for $\sigma$ above explicitly demonstrates that, even if the 
D-term error term is negligible, the uncertainty in fluxes in regions of source
emission is somewhat higher than the map rms in regions
far from source emission. 

In our case, $N_{ant} = 10$, $N_{IF} = 4$ for the 
7.9--15.4-GHz observations and 2 for the 4.6~GHz, 5.1~GHz and 1.36--1.67~GHz 
observations, and 
$N_{scan}\simeq 8$.  We estimate $\sigma_{\Delta}\simeq 0.005$ for all the
experiments from the scatter 
in the D-terms. The largest value for $\sigma_{Dterm}$ will occur at the peaks
of the maps; 
at the positions where we have determined the RM values below (Table~2),
the resulting D-term uncertainties are no more than $\simeq 0.60\sigma_{rms}$ 
for 0716+714 and no more than $\simeq 0.40\sigma_{rms}$ for 1749+701, making 
$\sigma_{Dterm}$ small compared to the other terms 
contributing to $\sigma$. 

The $Q$ and $U$ uncertainties determined in this way can then be propagated 
to derive the corresponding  uncertainties in the polarization angles, 
$\sigma_{\chi}$: 

\begin{eqnarray}
\chi & = & \frac{1}{2}ArcTan(\frac{U}{Q})\\
\sigma^{2}_{\chi} & = & \frac{1}{4}[(\frac{Q}{Q^{2}+U^{2}})^{2}\sigma_{U}^{2}
+(\frac{U}{Q^{2}+U^{2}})^{2}\sigma_{Q}^{2}]
\end{eqnarray}

\noindent
The uncertainty in the EVPA calibration $\sigma_{EVPA}$ can then
be added in quadrature:
\begin{equation}
\sigma^{2}_{\chi_{final}}=\sigma^{2}_{\chi}+\sigma^{2}_{EVPA}
\end{equation}

\noindent
These $\chi$ uncertainties can then, in turn, be used to determine 
uncertainties in the fitted RM values, as is described by Hovatta et al. 
(2012).  

Note, however, that, as the same EVPA calibration is applied to each 
polarization 
angle at a given frequency, the uncertainty this introduces is 
{\em systematic}. One consequence of this is that,  although the EVPA 
calibration uncertainties will increase the uncertainties in the 
fitted RM values, EVPA calibration uncertainties should not give rise to 
spurious RM gradients (e.g. Mahmud et al. 2009, Hovatta et al. 2012).  
The reason for this is 
essentially that any EVPA calibration error corresponds to 
a specific systematic offset that affects all EVPA measurements at all points 
of the maps at the corresponding frequency equally \emph{and in the same 
direction}, and so will not induce gradients between points.  

This was taken into account in our analysis in the same way as was done by
Mahmud et al. (2009) and Hovatta et al. (2012): when RM values are derived
specifically so that they can be compared to search for possible gradients,
the value $\sigma_{\chi}$ without adding $\sigma_{EVPA}$ in quadrature was
used to determine the RM uncertainty.

\section{Results}
\begin{figure*}
 \begin{minipage}[t]{7.5cm}
 \begin{center}
 \includegraphics[width=5.5cm,clip]{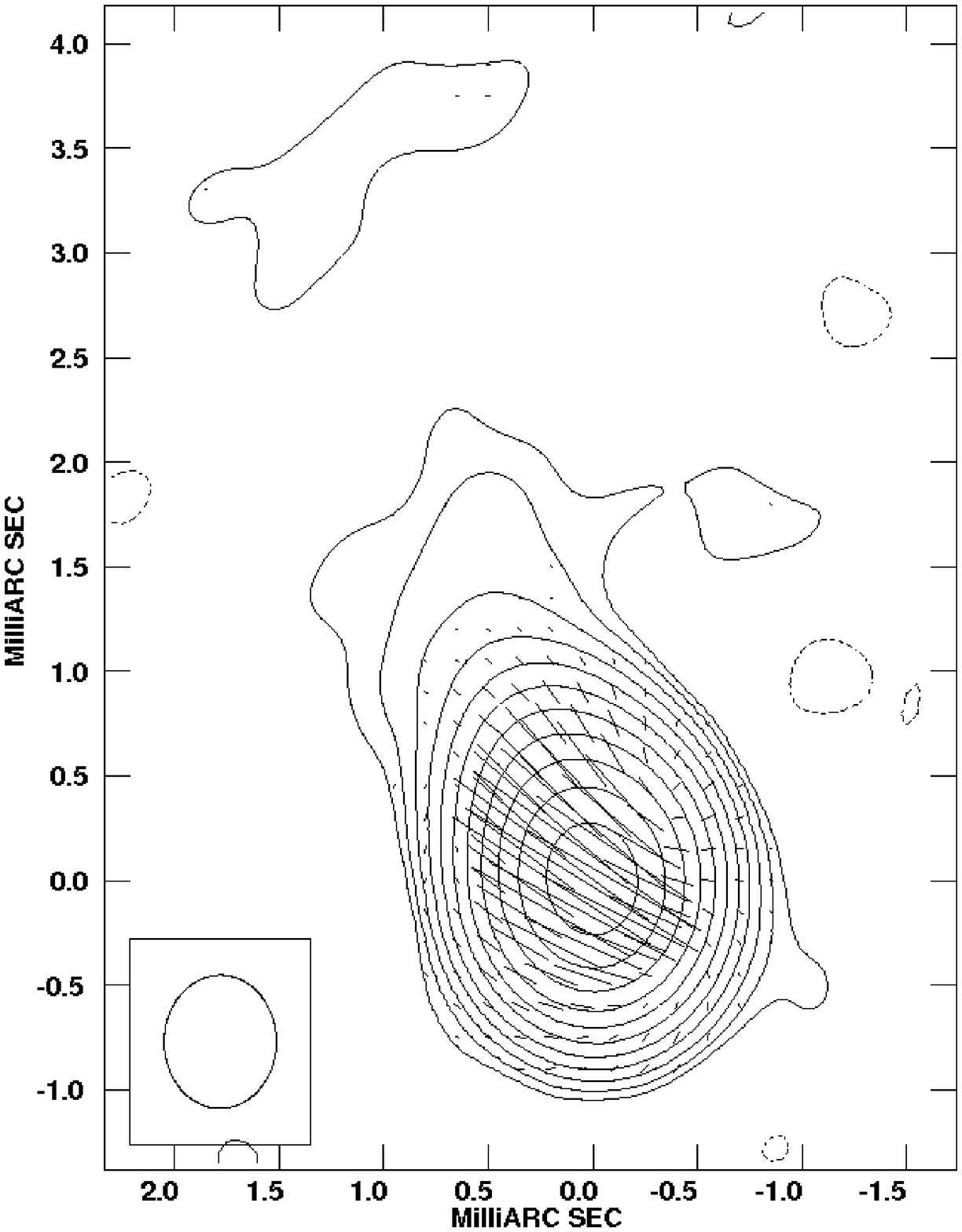}
 \end{center}
 \end{minipage}
 \begin{minipage}[t]{7.5cm}
 \begin{center}
 \includegraphics[width=7.0cm,clip]{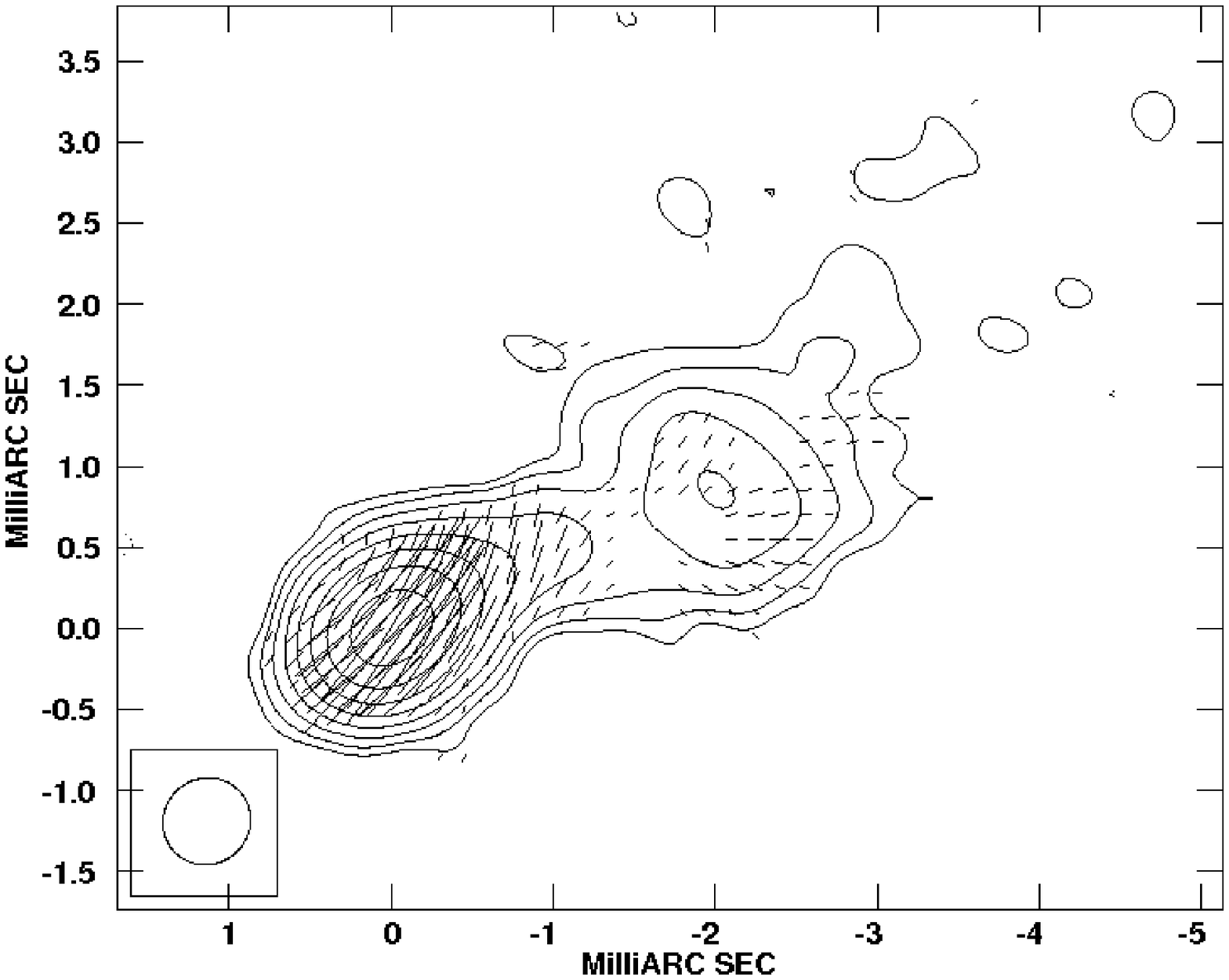}
 \end{center}
 \end{minipage}
\begin{minipage}[t]{7.5cm}
 \begin{center}
 \includegraphics[width=5.5cm,clip]{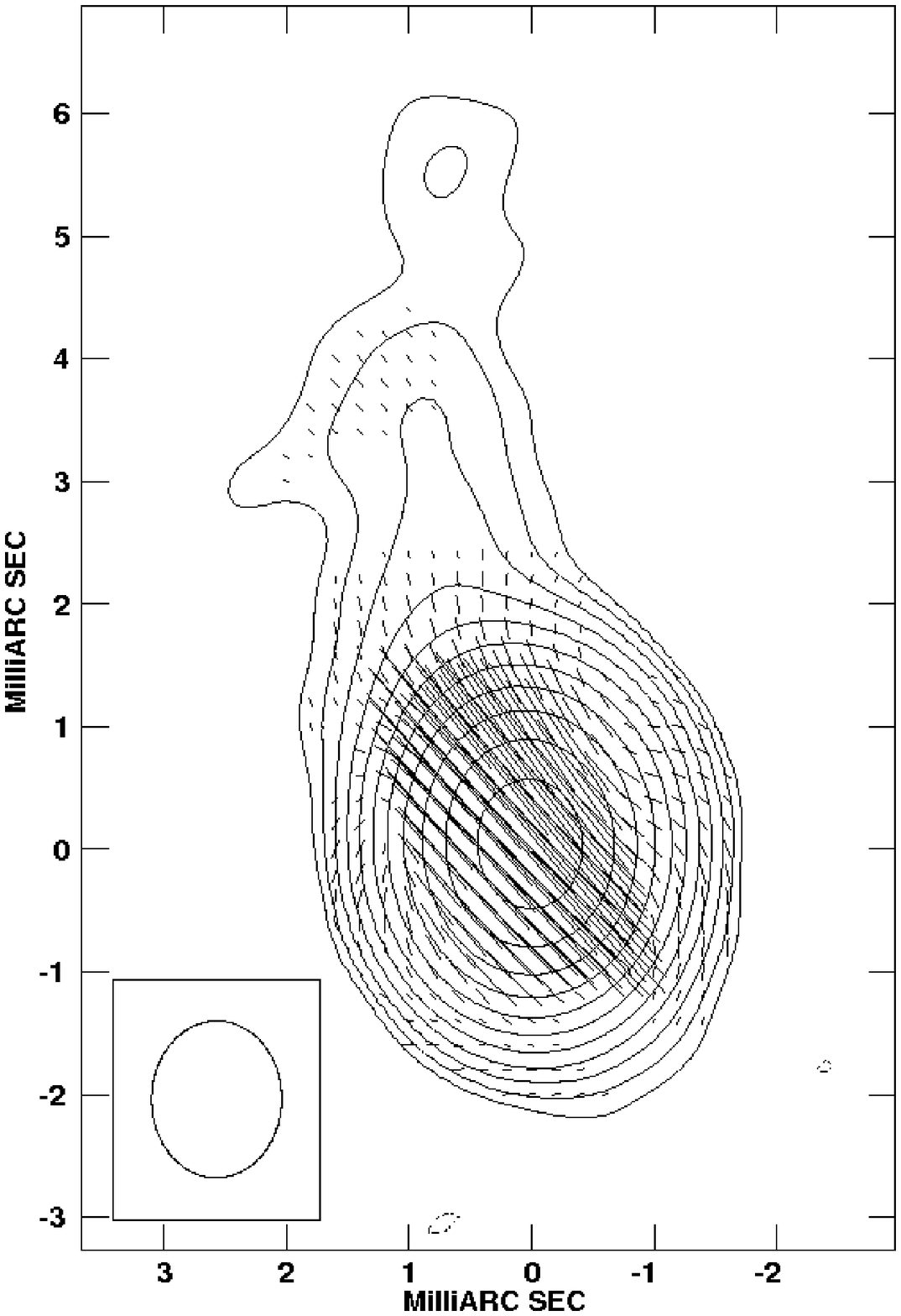}
 \end{center}
\end{minipage}
\begin{minipage}[t]{7.5cm}
 \begin{center}
 \includegraphics[width=7.0cm,clip]{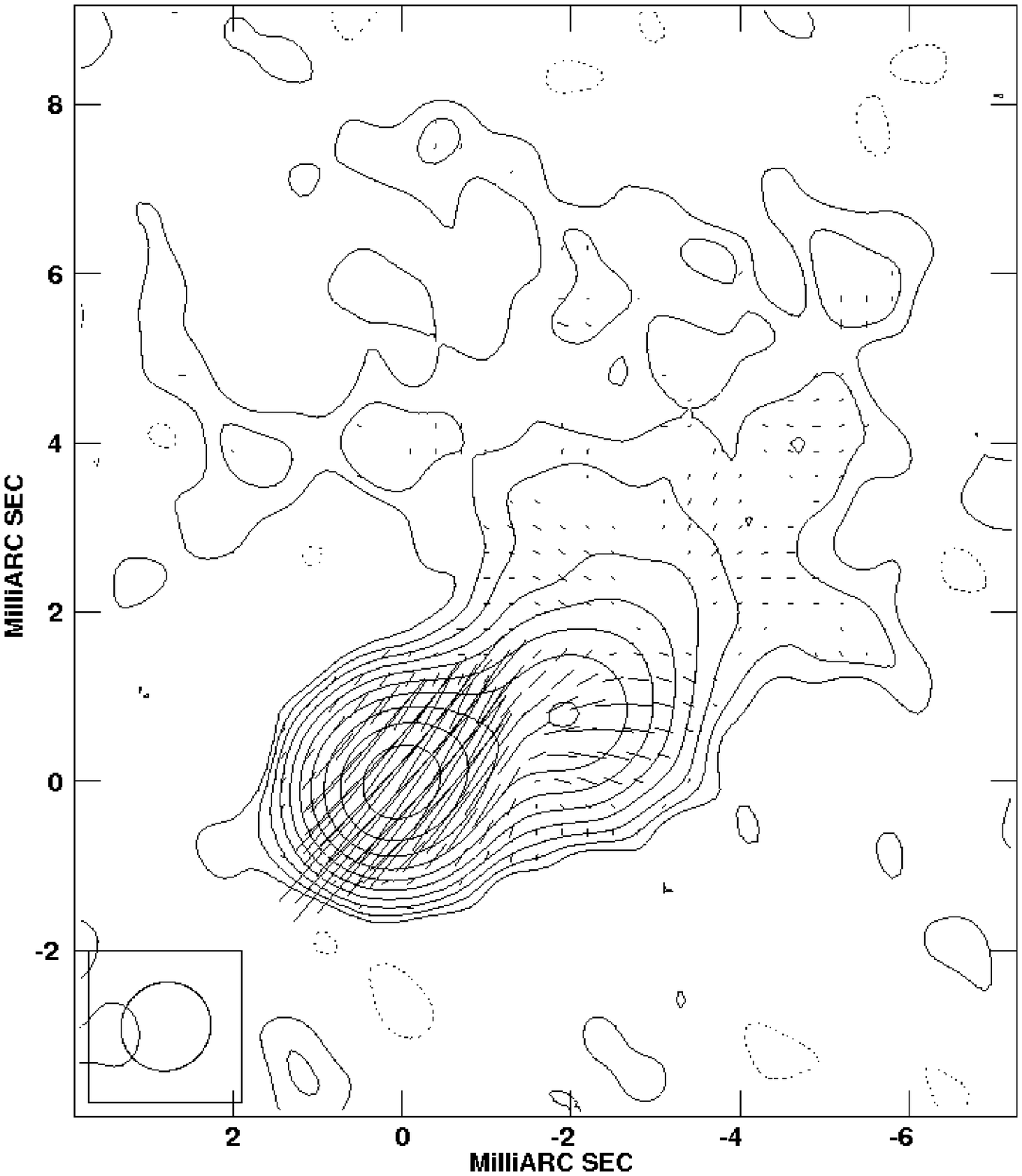}
 \end{center}
\end{minipage}
\begin{minipage}[t]{7.5cm}
 \begin{center}
 \includegraphics[width=5.5cm,clip]{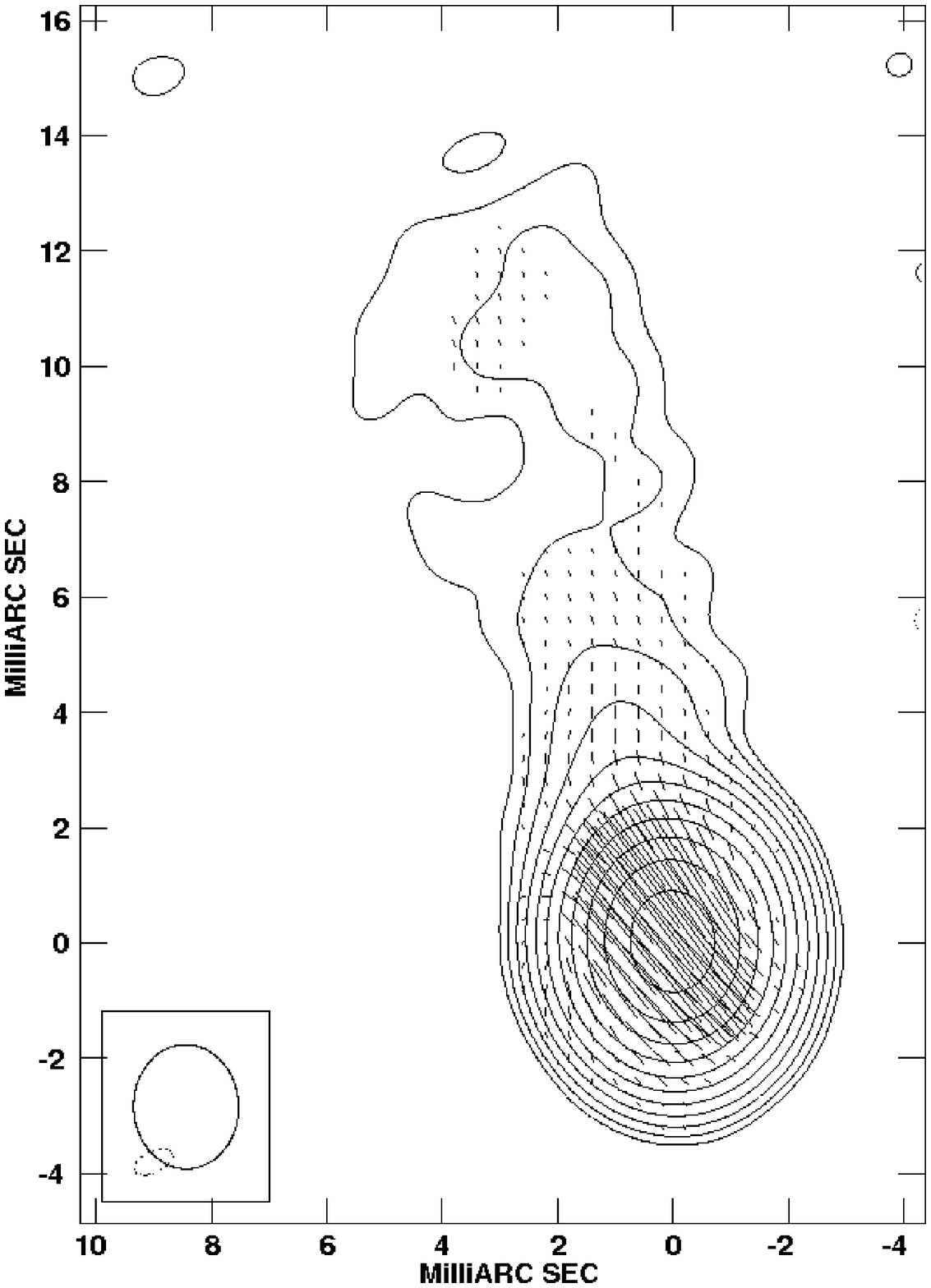}
 \end{center}
\end{minipage}
\begin{minipage}[t]{7.5cm}
 \begin{center}
 \includegraphics[width=7.0cm,clip]{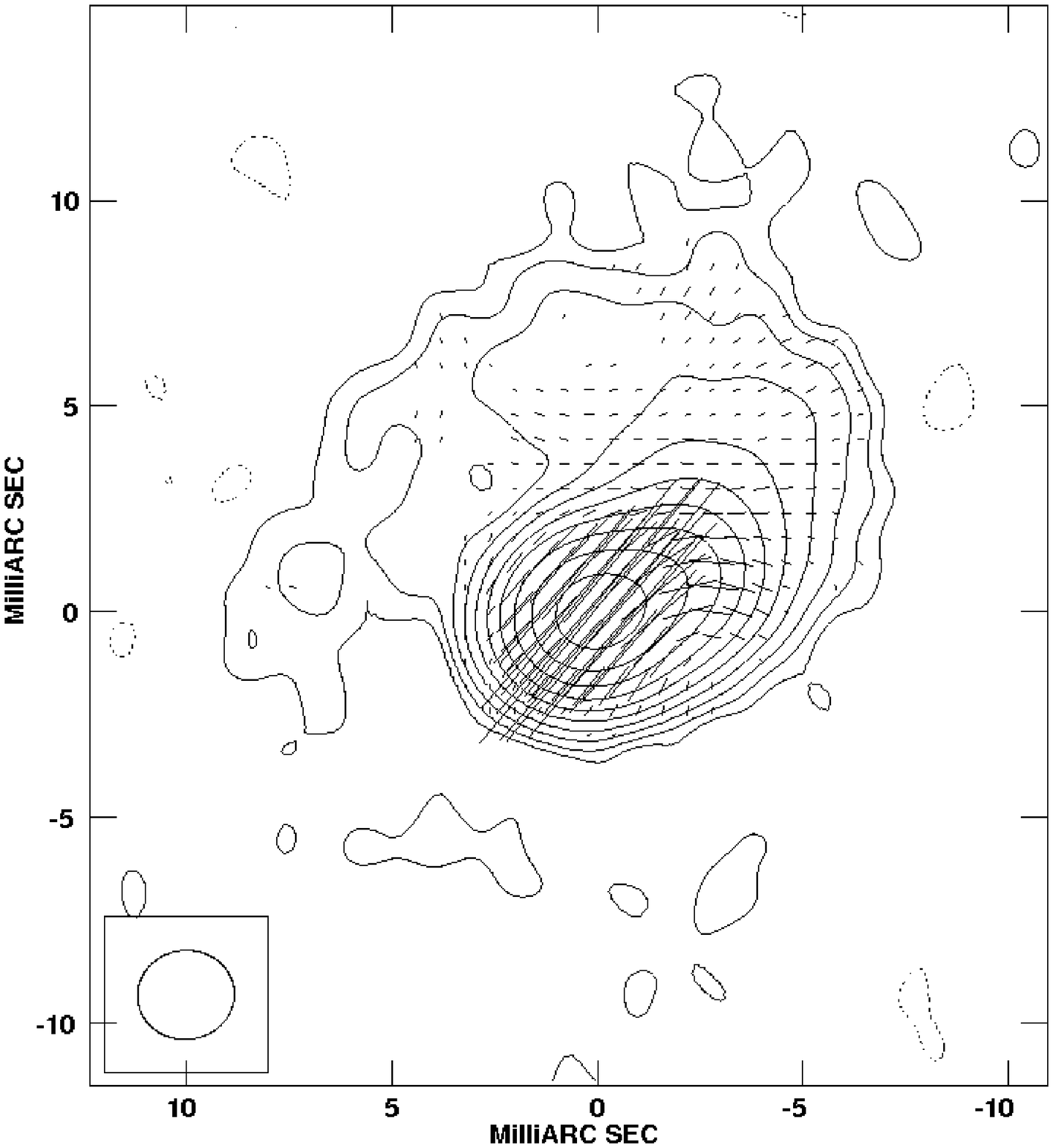}
 \end{center}
\end{minipage}
\caption[Short caption for figure 1]{\label{fig:pol_maps1}  VLBA $I$ maps 
for  0716+714 (left) and 1749+701 (right) with polarization sticks 
superimposed, at 15.4~GHz (top), 7.9~GHz (middle) and 4.6 GHz (bottom), 
corrected for integrated Faraday Rotation. The maps of 0716+714 have
peaks of 1.0, 1.3, and 1.5 Jy/beam and bottom contours of 0.7, 0.8 and
0.9 mJy/beam; the maps of 1749+701 have peaks of 0.5, 0.4 and 0.3~Jy/beam
and bottom contours of 0.6m, 0.5 and 1.7~mJy/beam. }
\end{figure*}

\begin{figure}
\centering
\includegraphics[width=0.5\textwidth]{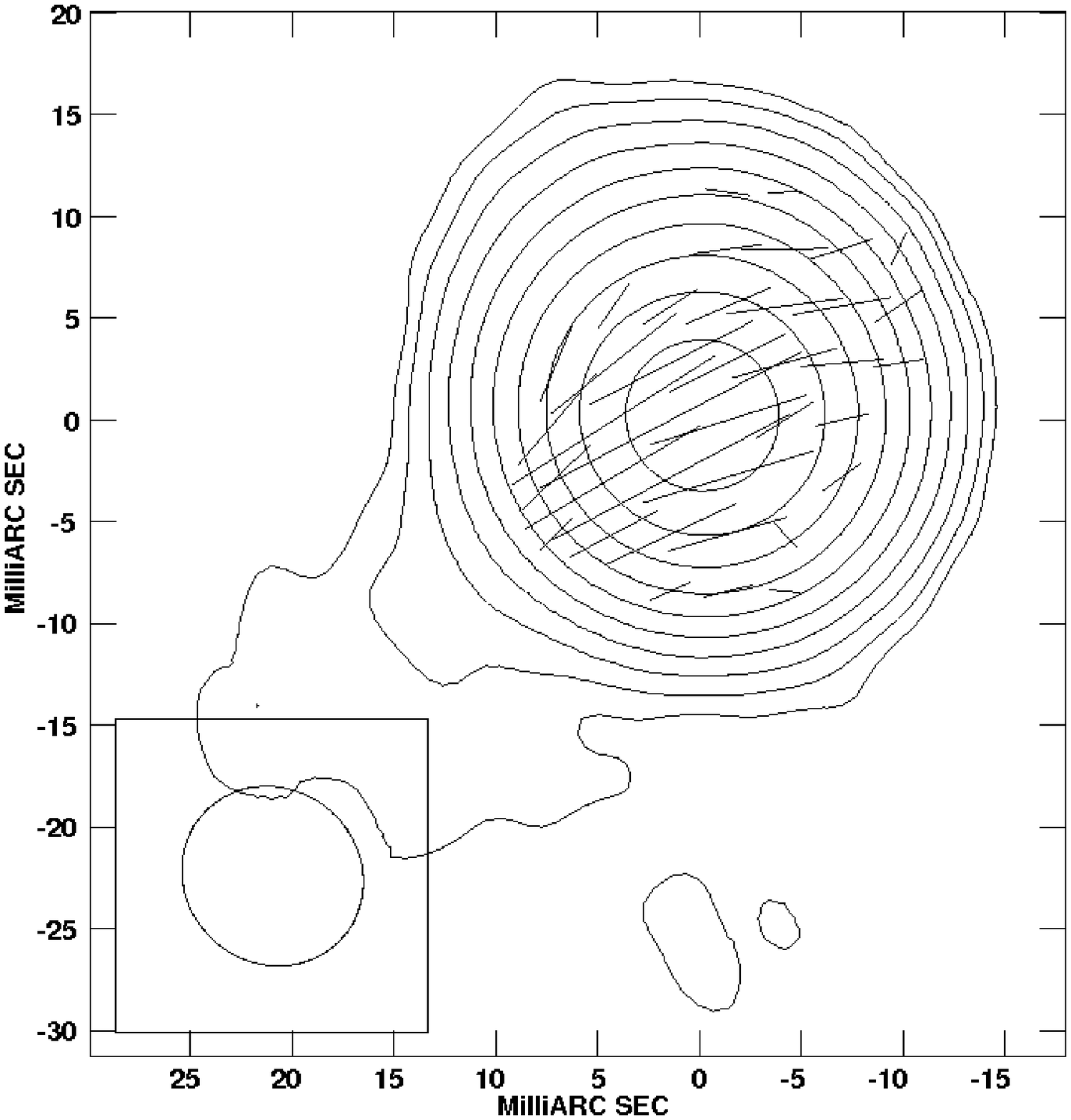}
\caption[Short caption for figure]{\label{fig:1749_pol_red}VLBA $I$ map 
of 1749+701 with polarization sticks superimposed, at 1.43~GHz, corrected 
for integrated Faraday Rotation.  The $I$ peak is 0.60~Jy/beam and the bottom
contour is 1.5~mJy/beam.}
\end{figure}

The images in Fig.\ \ref{fig:pol_maps1} show the observed VLBI 
total-intensity and linear-polarization structures for both sources at
15.4, 7.9 and 4.9 GHz, 
and Fig.~\ref{fig:1749_pol_red} the total-intensity and linear-polarization
structure of 1749+701 at 1.43~GHz, all corrected for integrated but not 
local Faraday rotation. The maps at 1.36, 1.49 and 1.67~GHz are very similar,
and are not shown here. The convolving beams used in each case are indicated 
in the lower right-hand corner of the figures. The peaks and bottom contours
are indicated in the figure captions and, in all cases, the contour levels
increase in steps of a factor of two.

0716+714 has a redshift of $z = 0.30$, corresponding
to 4.52 pc/mas, and an integrated RM of $-30$~rad/m$^2$ (Pushkarev 2001).
The jet of 0716+714 extends roughly to the North. The 
jet polarization {\bf E} vectors are aligned with the local jet direction, 
as is also shown by the 2cm MOJAVE images (http://www.physics.purdue.edu/MOJAVE/sourcepages/). 

1749+701 has a redshift of $z = 0.77$, corresponding 
to 7.41~pc/mas, and an integrated RM of $+15$~rad/m$^2$ (Pushkarev 2001).
The jet of 1749+701  
initially emerges toward the Northwest, then turns toward the North, and 
further toward the East; this spiral-like path is evident in the 7.9 and 4.9-GHz 
maps (see also Gabuzda \& Lisakov 2009). The polarization {\bf E} 
vectors are mostly aligned with the local jet direction; some regions of 
`spine-sheath' polarization structure or orthogonal polarization offset
toward one side of the jet are visible in the maps. Although the core--jet 
structure is not directly distinguishable in the 1.43-GHz image for 
January 2004 in Fig.~\ref{fig:1749_pol_red}, the orientation of this structure 
is known from the higher-frequency images in Fig.~\ref{fig:pol_maps1}. 
The polarization {\bf E} vectors appear to be aligned with the 
jet direction. The weak emission to the southeast of the map center corresponds
to a continuation of the emission in roughly this region in the 4.9-GHz image
(Fig.~\ref{fig:pol_maps1}, bottom right).

The images in Figs.~\ref{fig:0716_RM} and \ref{fig:1749_RM} show the 
parsec-scale RM distributions for 0716+714 (4.6--15.4~GHz) and 1749+701 
(1.36--1.67~GHz), superimposed on the corresponding $I$ contours. The RM distribution for
1749+701 for 4.6--15.4~GHz is subject to uncertainty due to the relatively large
shifts required to align the $\chi$ images at the different frequencies, and
we accordingly focus on the more reliable images shown in these two figures.
In all cases, the $I$ contours increase in steps of a factor 
of two. The arrows show the direction of RM gradients in the corresponding 
regions visible by eye; in other words, the direction in which the value 
of the RM increases (from more negative to less negative, negative to positive,
or less positive to more positive, as the case may be). 
The convolving beams used in each case are indicated in the lower 
right-hand corner of the figures. The beam used for the 4.6--15.4~GHz RM map
for 0716+714 was $1.28$~mas~$\times 1.06$~mas in position angle $-0.84^{\circ}$,
which corresponds to the resolution of the 7.9~GHz data; this beam was chosen
in order to provide slightly higher resolution in the RM map at the expense 
of only a modest over-resolution of the lowest-frequency images.
Accompanying panels show plots of 
polarization angle ($\chi$) vs. wavelength squared ($\lambda^2$) for the 
indicated regions; the uncertainties in the polarization angles shown here 
include the EVPA uncertainty added in quadrature. 
Slices of the RM across the gradients in the specified locations 
in the jets and core regions obtained with the AIPS task 'SLICE' are also
shown; we do not include the (single-pixel-based) uncertainties on these
slices on these plots, since they are meant only to be orientational. Instead,
we carry out below an anaysis involving the RM values and their 
uncertainties for three regions
across the jet (on either side and in the center), more in keeping with the
limited resolution available with our arrays.

\section{Discussion}
 
\subsection{Linear Polarization Structure}
Previous polarization observations have demonstrated that `spine-sheath' 
polarization structures are not uncommon among blazars. Attridge et al. 
(1999) interpreted a `spine-sheath' polarization structure in the jet 
of the quasar 1055+018 as a result of a series of shocks compressing the field 
in the central region and shearing of the field induced by interaction with 
the surrounding medium at the jet edges. However, more recent studies 
(eg. Lyutikov et al. 2005, Pushkarev et al. 2005) have discussed the 
possibility that `spine-sheath' polarization structures can come about 
naturally in the case of helical jet magnetic fields. The BL~Lac objects 
0714+716 and 1749+701 both show signs of `spine-sheath' polarization or 
transverse polarization offset toward one side of the jet at one or more 
wavelengths, consistent with the
possibility that their jets carry helical magnetic fields. 

\subsection{Detection of Transverse RM Gradients across the Jets}
\begin{figure*}
\begin{minipage}[t]{16.0cm}
   \begin{center}
  \includegraphics[width=16.0 cm,clip]{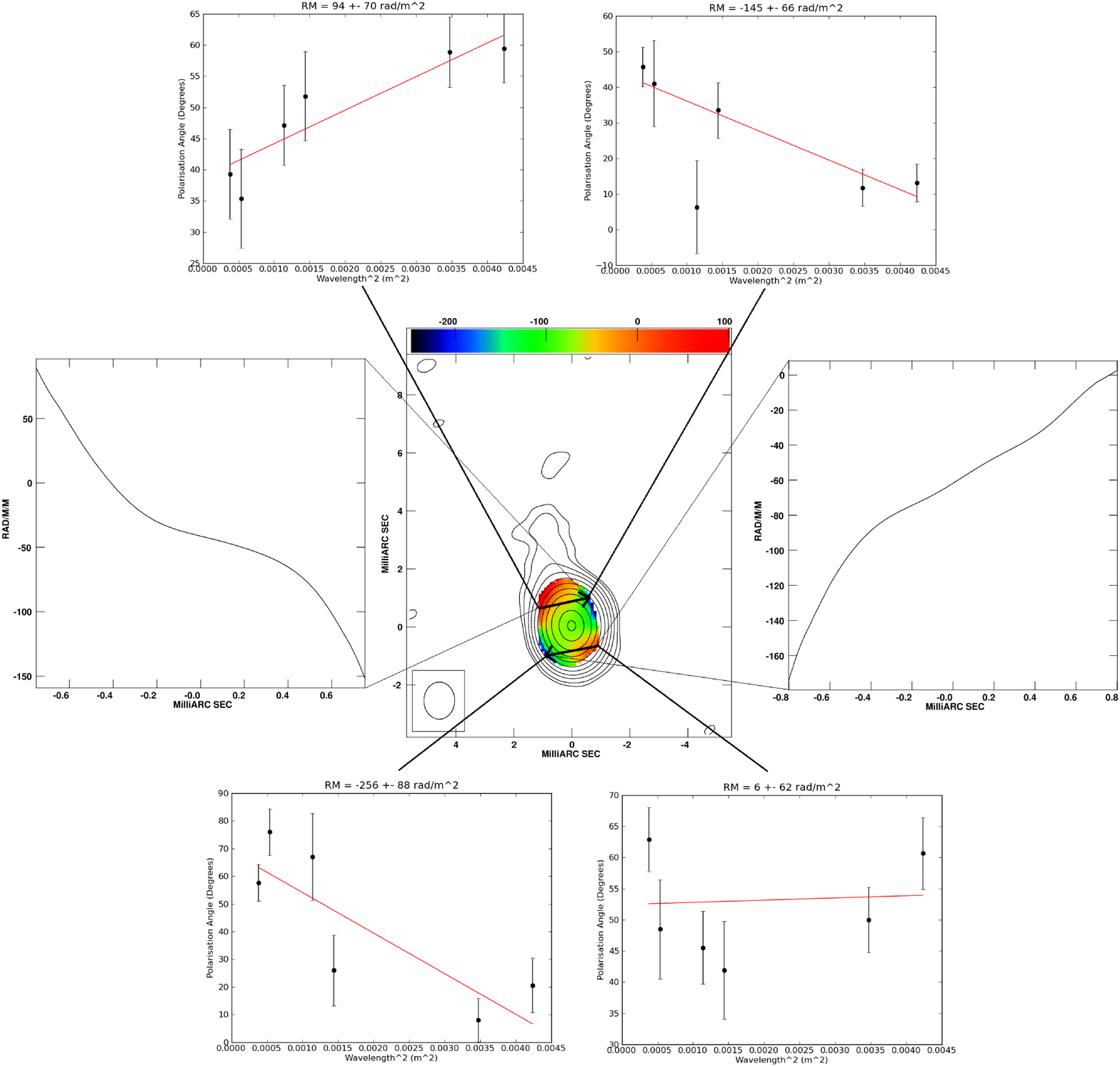}
 \end{center}
   \end{minipage}
\caption{\label{fig:0716_RM} RM map of 0716+714 at 4.6--15.4~GHz.
The accompanying panels show slices 
of the RM distributions across the jet and core, and polarization angle 
($\chi$) vs. wavelength-squared ($\lambda^2$) plots for pixels on
either side of the core and jet. The errors shown in the plots are 1$\sigma$,
and include the estimated random errors and the EVPA uncertainties added in
quadrature. 
The peak of the $I$ map is 1.3~Jy/beam and the bottom 
contour is 1.0~mJy/beam. The beam used to construct the $I$ and RM 
maps was 1.28~X~1.06~mas in position angle $-0.8^{\circ}$.}
 \end{figure*}

\begin{figure*}
\centering
\includegraphics[width=0.6\textwidth,angle=90]{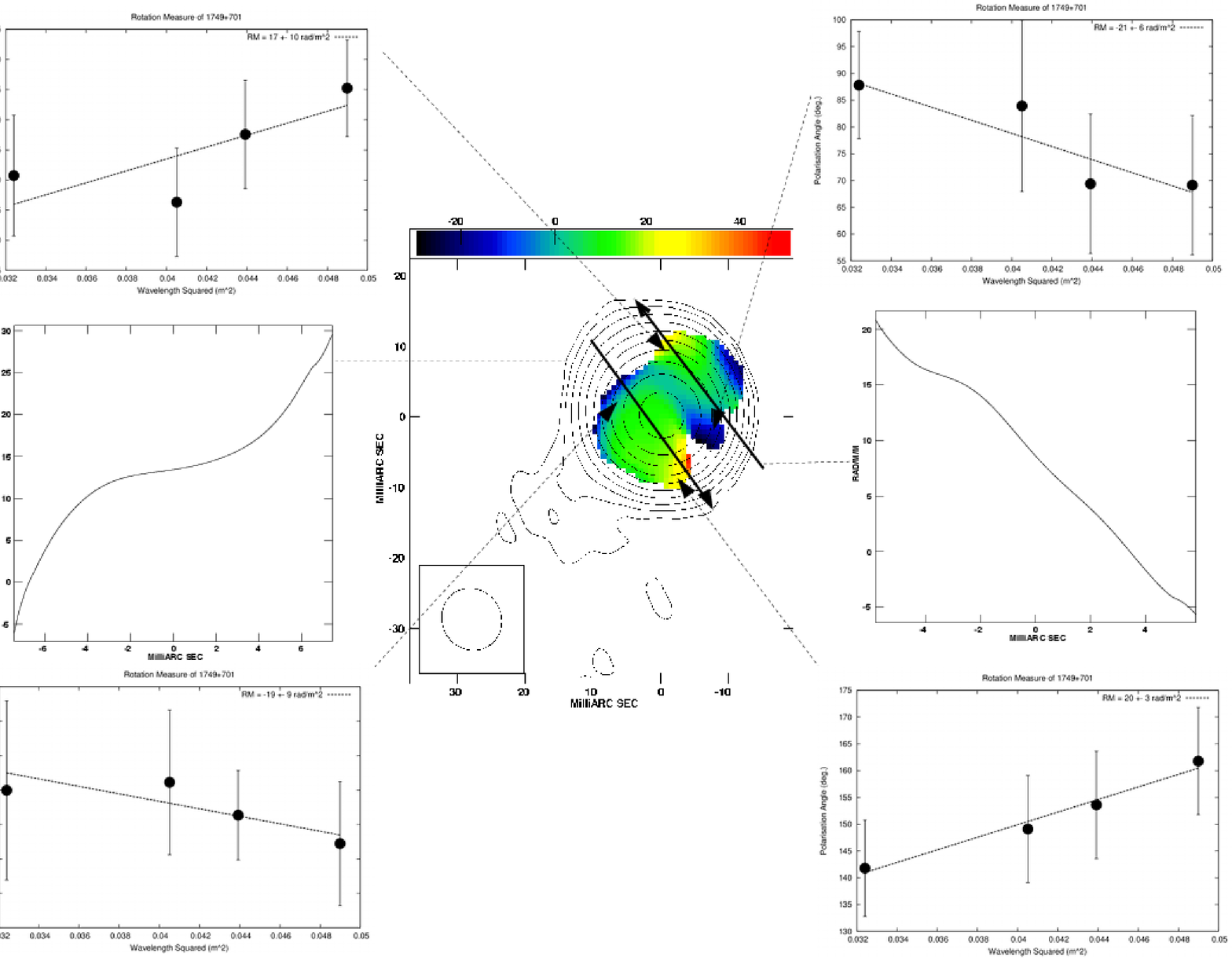}
\caption{\label{fig:1749_RM} RM map of 1749+701  
1.36--1.66~GHz. 
The accompanying panels show slices of the RM distribution 
across the core, and polarization angle ($\chi$) vs. wavelength-squared 
($\lambda^2$) plots for pixels on either side of the core and
jet.  Errors shown are 1$\sigma$, and include the estimated random errors and
the EVPA uncertainties added in quadrature. 
The peak of the $I$ map is 
0.6~Jy/beam; the bottom contour 
is 1.4~mJy/beam (January 2004).
The beam used to construct the $I$ and RM maps was 9.16~X~8.57~mas
in position angle $49^{\circ}$.}
\end{figure*}

Tentative transverse RM gradients across the jets of the two BL~Lac 
objects considered here are visible by eye in the colour versions of the RM
distributions in Figs.~\ref{fig:0716_RM} and \ref{fig:1749_RM}. 
The first step in testing the reality of these gradients is estimating
the uncertainties in the RM values on either end of the gradient, to 
determine at what level the two RM values differ. We have done this 
using the $Q$ and $U$ uncertainty relations of Hovatta et al. (2012), 
as described above.

\begin{figure*}
\centering
 \includegraphics[width=0.33\textwidth,angle=-90]{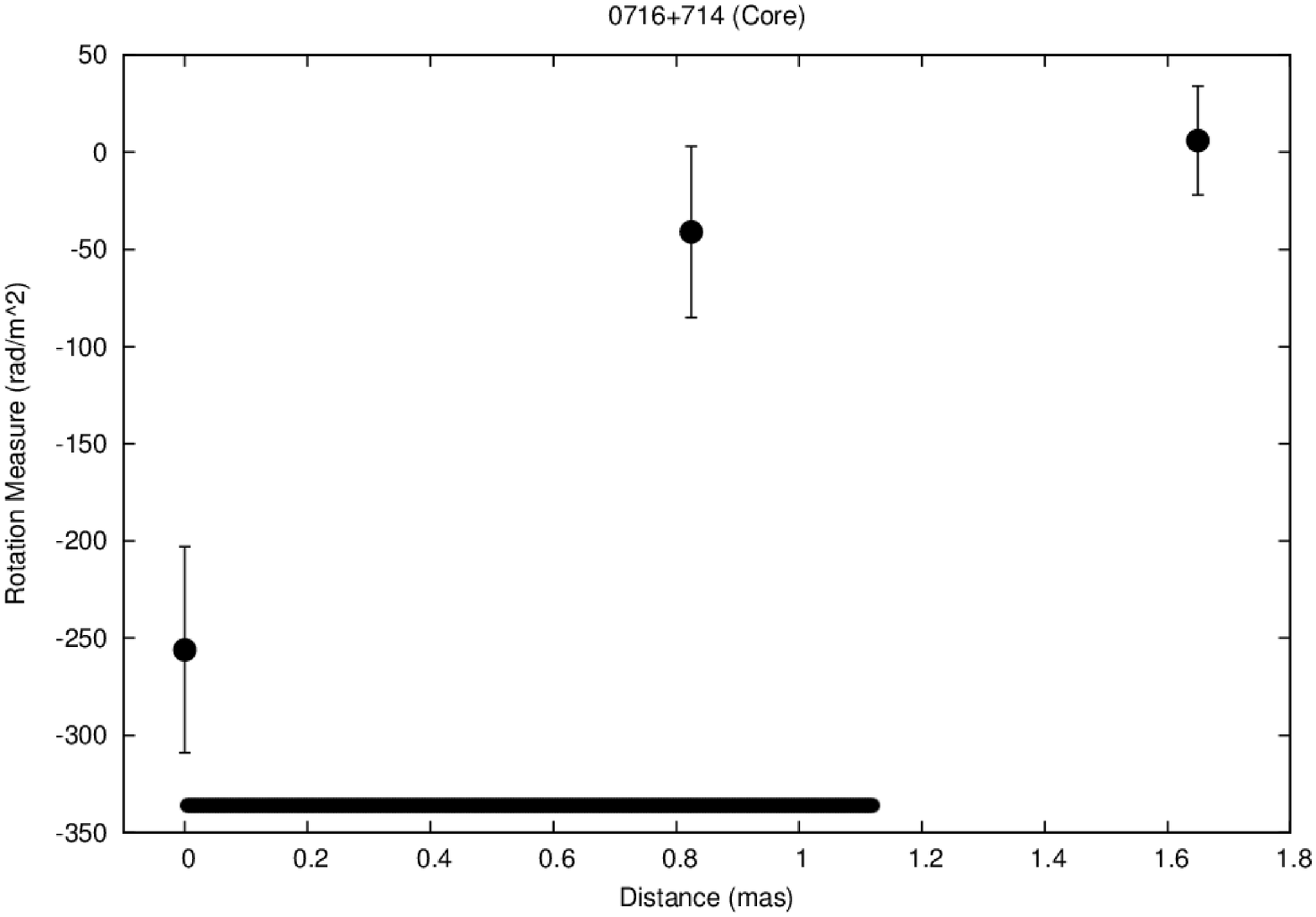}
 \includegraphics[width=0.33\textwidth,angle=-90]{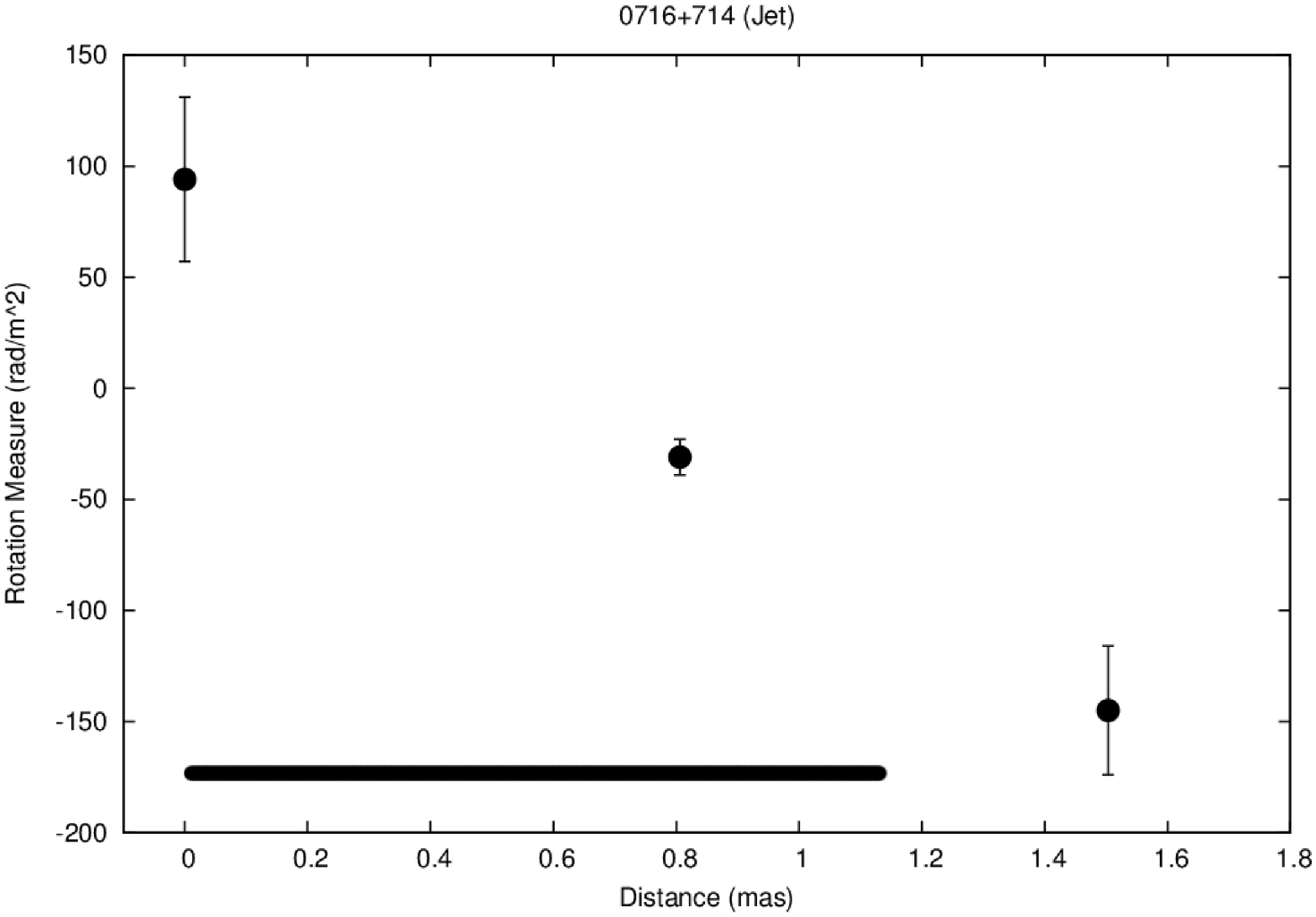}
 \caption[Short caption for figure]{\label{fig:0716_transdist} Plots of 
observed RM as a function of distance from a reference point 
on one side of the source structure across the core-region (left) and jet 
(right) RM distributions of 0716+714 at 4.6--15.4~GHz. 
The positions of each point and the corresponding RM values and their errors
are listed in Table~2. The horizontal bar shows the approximate size of the
beam FWHM.}
\end{figure*}

\begin{figure*}
\centering
\includegraphics[width=0.33\textwidth,angle=-90]{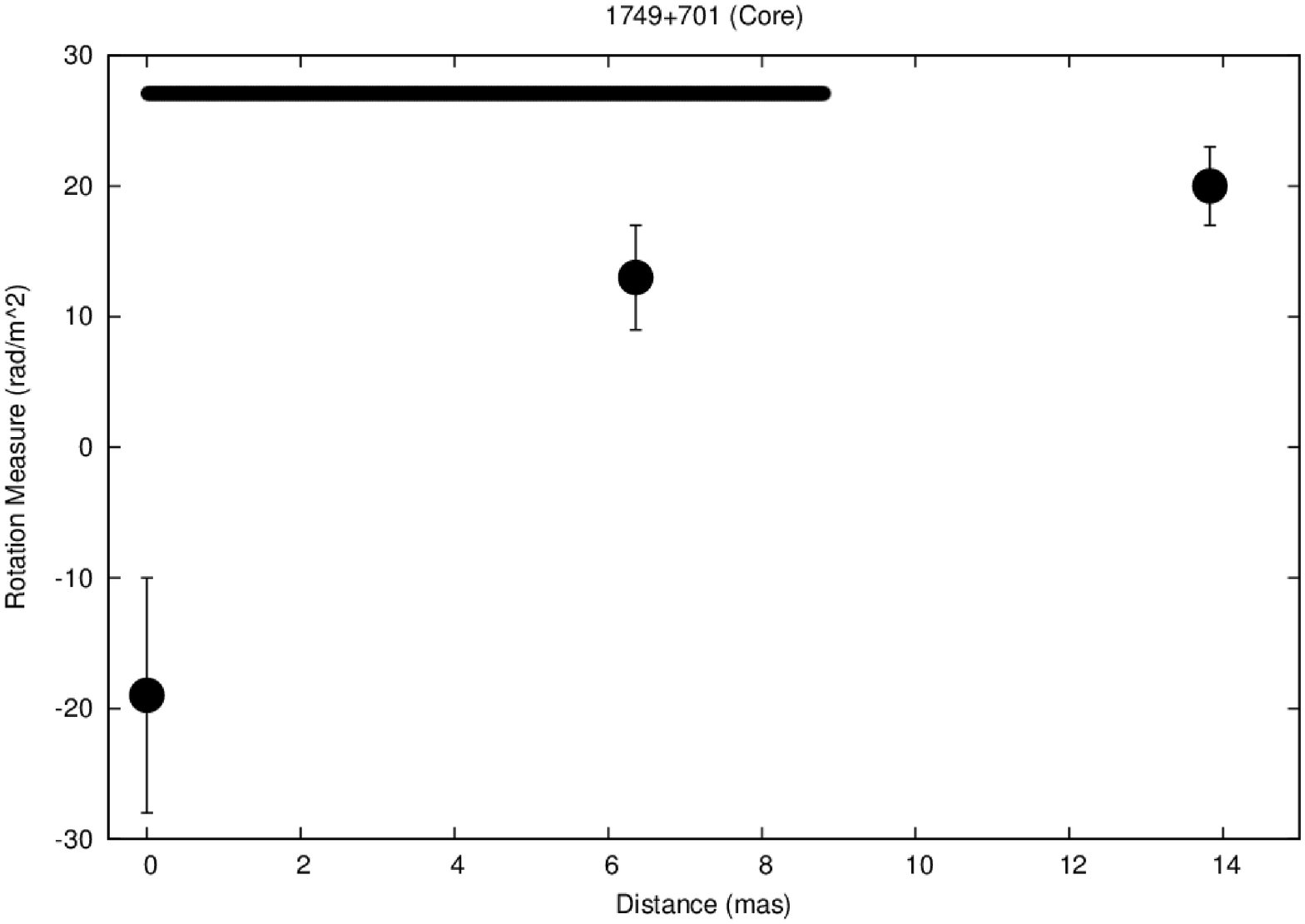}
\includegraphics[width=0.33\textwidth,angle=-90]{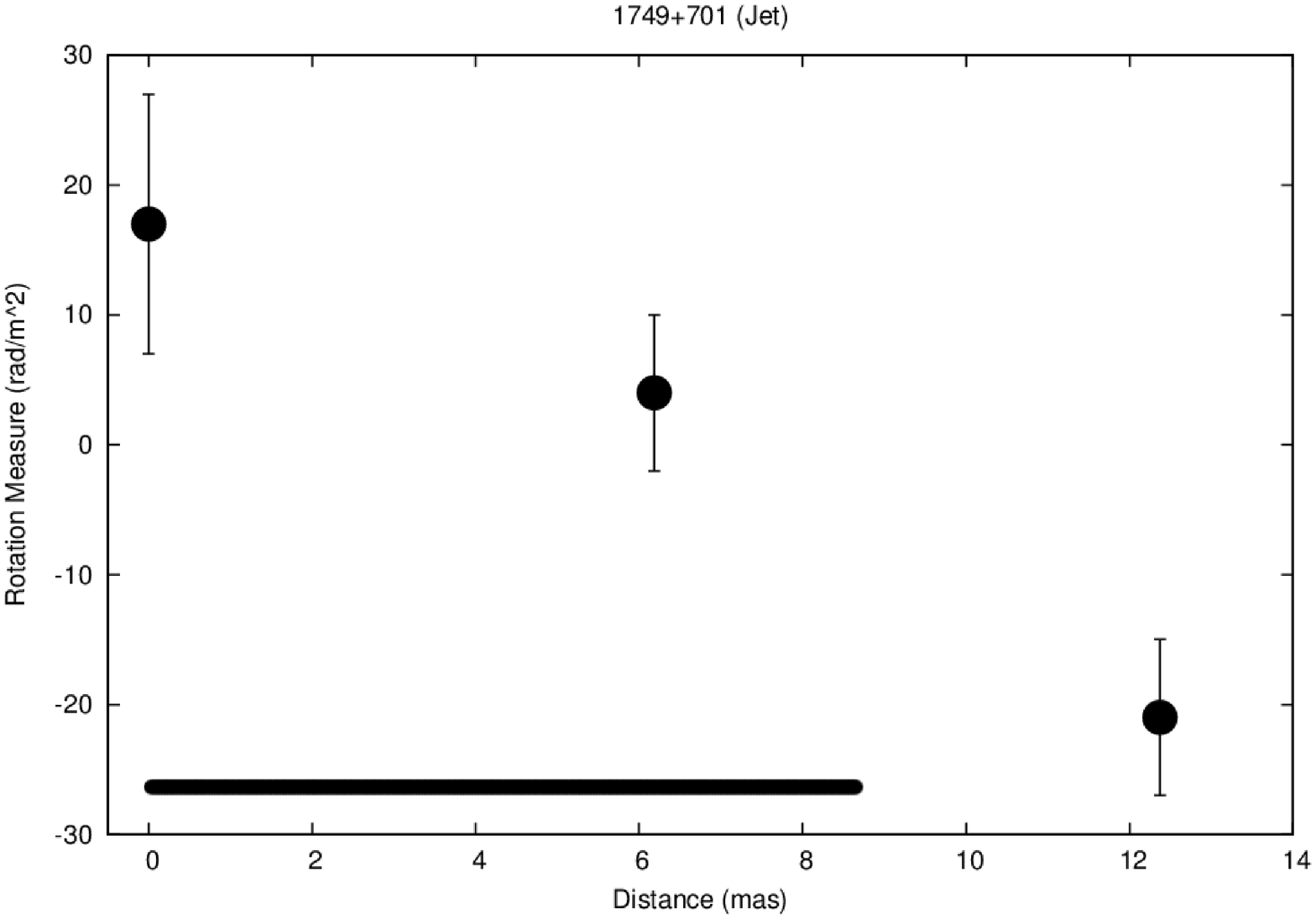}
\caption[Short caption for figure]{\label{fig:1749_transdist} Plots of 
observed RM as a function of distance from a reference point            
on one side of the source structure across the core-region (left) and jet       
(right) RM distributions of 1749+701 at 1.36--1.67~GHz. 
The positions of each point and the 
corresponding RM values and their errors are listed in Table~2. The 
horizontal bar shows the approximate size of the beam FWHM.}
\end{figure*}

Figs.~\ref{fig:0716_transdist} and \ref{fig:1749_transdist} show plots of 
the observed RMs at three points across the 
core-region and jet structures of 0716+714 and 1749+701 (on either side
and near the center).  Together with the slices shows in
Figs.~\ref{fig:0716_RM} and \ref{fig:1749_RM}, these figures demonstrate
the systematic, monotonic nature of these observed RM gradients.

Table~2 summarizes the sets of RM values shown in Figs.~\ref{fig:0716_transdist} 
and \ref{fig:1749_transdist} together 
with their uncertainties, as well as the differences between the  RM values 
on either side of the inferred transverse gradients and their uncertainties. 
The columns present (1) the figure to which the RM values refer, (2) the
source name, (3) the point in the indicated figure to which the RM
value corresponds, (4) the position where the RM value was measured, in
milliarcseconds, relative to the phase centre, (5) the RM value 
at the indicated position, together with its
uncertainty, (6) the difference between the two RM 
values on either side of the source structure and its uncertainty
and (7) the significance of 
this difference in numbers of $\sigma$. The uncertainties listed in 
column (5) are based on $\chi$ uncertainties without the EVPA-calibration
uncertainty added in quadrature, since this will affect all points in an
RM image systematically in the same way.
The last column of this table shows that the differences between the RM
values detected on either side of the jet structures are at the level
of $3-5\sigma$, demonstrating that these differences appear to be 
statistically significant. 

\begin{table*}
\caption{RM measurements in Figs. \ref{fig:0716_transdist} and \ref{fig:1749_transdist}} \centering
\begin{tabular}{llcccccccc}
\hline
Figure  & Source   & Point  & Position         & RM          & Left--Right & RM Diff in $\sigma$\\
        &          & in plot&  (mas)           &rad/m$^2$    & RM Diff &\\ 
(1)     & (2)      & (3)    & (4)              & (5)         & (6)     & (7) \\ \hline
\ref{fig:0716_transdist} (left)& 0716+714 & Left   & $(+0.90, -1.00)$ & $-256\pm 53$& &\\
        &  (Core-region)  & Mid    & $(0.10, -0.80)$ & $-41\pm 44$ & $+262\pm60$ & $4.4\sigma$\\
        &          & Right  & $(-0.70, -0.60)$ & $+6\pm 28$ & &\\
\ref{fig:0716_transdist} (right)& 0716+714& Left   & $(+1.00, +1.00)$ & $+94\pm 37$ & &  \\
         &  (Jet)  & Mid    & $(+0.30, +1.00)$ & $-31\pm 9$ & $-239\pm 47$& $5.1\sigma$\\
         &         & Right  & $(-0.50, +1.10)$ & $-145\pm 29$& &\\
\ref{fig:1749_transdist} (left) & 1749+701& Left   & $(+9.00, +1.50)$ & $-19\pm 8$ & & \\
         & (Core-region)  & Middle & $(+4.50, -3.00)$ & $+13\pm 4$ & $+39\pm 9$    & $4.3\sigma$\\
         &         & Right  & $(-1.50, -7.50)$ & $+20\pm 2$ &&\\
\ref{fig:1749_transdist} (right)& 1749+701& Left   & $(-4.50, +9.00)$ & $+17\pm 10$ &&\\
         &  (Jet)  & Middle & $(-6.00, +3.00)$ & $+4\pm 6$ & $-38\pm 12$   & $3.2\sigma$\\
         &         & Right  & $(-7.50, -3.00)$ & $-21\pm 6$ &&\\
\hline
\end{tabular}
\end{table*}

\subsection{The Possibility of Detecting Transverse RM Gradients Across
Narrow Jets}

Taylor \& Zavala (2010) have recently proposed four criteria for the
reliable detection of transverse Faraday rotation gradients, the most
stringent of which is that the observed RM gradient span at least three 
``resolution elements'' across the jet. This criterion reflects the desire 
to ensure that it is possible to distinguish properties between regions 
located on opposite sides of the jets. The criterion of three ``resolution
elements'' has been taken to correspond to three beamwidths, and coincides
with the general idea that structures separated by less than a beamwidth
are not well resolved.

To test the validity of this criterion of Taylor \& Zavala (2010), we 
constructed core--jet-like sources with various intrinsic widths and with 
transverse RM gradients present across their structures, and carried out Monte 
Carlo simulations based on these model sources. 
A  description of our Monte Carlo simulations and the results they
yielded are presented in the Appendix.  The transverse source widths for 
our model sources correspond to intrinsic widths of about 1/2, 1/3, 1/5, 1/10
and 1/20 of the beam full-width at half-maximum (FWHM) in the direction 
across the jet.  The simulations show that the transverse 
RM gradients introduced into the model visibility data remain visible 
in the RM maps constructed from ``noisy'' data having the same distribution 
of $(u,v)$ points as our observations of 0716+714, even when the intrinsic 
width of the structure is much smaller than the beam width. Both uni-directional 
model RM gradients and model RM structure containing two oppositely directly 
transverse gradients in the core region and jet are visible for all 
the jet widths considered. 

The results of these new Monte Carlo simulations thus directly demonstrate that 
the three-beamwidth criterion of Taylor \& Zavala (2010) is overly restrictive,
since the simulations directly show the possibility of detecting transverse RM 
gradients even 
when the intrinsic widths of the corresponding source structures are much less 
than the beamwidth, resulting in RM distributions that span only $1-1.5$
beamwidths. This demonstrates that the relatively modest widths spanned by the
transverse RM gradients in 0716+714 and 1749+701 that we report here should not
be taken by themselves as grounds to question the reliability of these gradients.
We note here that our Monte Carlo simulations are not intended to provide a
physical model for our observations, or to reproduce our observed RM distributions
in any detail; instead, they are intended solely to demonstrate the possibility
of detecting a transverse RM gradient in real data, even if the intrinsic jet
width is much smaller than the beam FWHM.

Inspection of Fig.~30 of Hovatta et al. (2012) indicates that the 
fraction of ``false positives'', i.e., spurious RM gradients, that were obtained
in their Monte Carlo simulations did not exceed $\simeq 1\%$ when a $3\sigma$ 
criterion was imposed for the RM gradient, even when the observed width 
of the RM gradient was less than 1.5 beamwidths. This suggests that there may
be up to a $\simeq 1\%$ probability that the RM gradients we report here are 
spurious, due to their relatively limited widths, although we consider this
to be unlikely, given that the RM differences involved correspond to 
as much as $5\sigma$.  

\subsection{The Remaining Criteria of Taylor \& Zavala (2010)}

With regard to the other criteria for reliability of transverse RM gradients
proposed by Taylor \& Zavala (2010), the criterion that the change in the 
RM across the jet be at least $3\sigma$ is satisfied by the RM images in 
Figs.~\ref{fig:0716_RM} and \ref{fig:1749_RM} (see also Table~2). The 
differences in the RMs across the core region and jet of 0716+714 
(Fig.~\ref{fig:0716_transdist}) are approximately $4-5\sigma$; the differences 
in the RMs across the core region and jet of 1749+701 
(Fig.~\ref{fig:1749_transdist}) are approximately $3-4\sigma$. 

The criterion that the change in the RM be monotonic and smooth within the 
errors is also satisfied by the gradients in both 0716+714 and in 1749+701.
Although the gradients suggested by the slices displayed in Figs.~\ref{fig:0716_RM} 
and \ref{fig:1749_RM} 
are not constant (linear), they are nevertheless monotonic.  It is interesting
to note here that the simulated RM maps of Broderick \& McKinney (2010) typically 
do not show RM gradients with a constant slope all across the RM distribution
after convolution, even though the intrinsic predicted gradients are monotonic 
[see, for example, their Fig.~8 bottom right panels].

The remaining criterion proposed by Taylor \& Zavala (2010) is that the spectrum
be optically thin at the location of the observed RM gradient.  This criterion
is motivated by two factors: (i) the desire to avoid possible jumps in the observed
polarization angles due to optically thick--thin transitions with the observed
frequency range, and (ii) the fact that the fractional polarization can change
rapidly with optical depth in the optically thick regime, leading to the
possibility of wavelength-dependent polarization effects when regions having
different optical depths at different frequencies are superposed, which could in
principle lead to the fitting of spurious RM values in optically thick regions
when these are inhomogeneous. 

This criterion is clearly satisfied by the gradients across the jets of 
0716+714 and 1749+701, which are all optically thin. 
The core regions of these two objects are also predominantly, but not fully, 
optically thin.  The core-region spectral indices and $\chi$ values provide no evidence 
for a transition between optically thick and optically thin in the frequency 
rangees considered, consistent with the fact that the observed Faraday rotations 
in the polarization angles are all no greater than a few tens of degrees.
Thus, there is no reason to suspect that jumps in the observed polarization 
angles due to optical-depth transitions are contributing to the observed 
core-region RMs. We cannot completely rule out the possibility that
the polarization angles in the core-region are subject to wavelength-dependent
optical-depth effects, however, we consider this to be unlikely, for two reasons:
(i) the degrees of polarization in the core regions are $m_{core}\simeq 3-4\%$ 
for 0716+714 and $m_{core}\simeq 7\%$ for 1749+701, indicating a substantial 
contribution from optically thin regions; and (ii) the quality of the $\lambda^2$ 
fits for the core regions is no worse than for the optically thin jet regions. 

Thus, our detection of the RM gradients across the jets can be considered 
firm, while the detection of the oppositely directed RM gradients across the
core may be somewhat more tentative, due to the small possibility that
the observed polarization could be affected by optical depth effects at
some of the observed frequencies. This is much less likely to be the case for
1749+701, since the observed frequencies span the relatively narrow range from
1.36--1.67~GHz.

\subsection{Reversal of RM gradients in the Core Region and Jet}

In both 0716+714 and 1749+701, the tentative transverse RM gradients 
detected in the core region are opposite in direction to the RM gradients 
detected across the jets 
(Figs.~\ref{fig:0716_RM}--\ref{fig:1749_RM} and 
Figs.~\ref{fig:0716_transdist}--\ref{fig:1749_transdist}).
In fact, a similar behaviour is
shown by the parsec-scale RM distribution for 3C~120 presented by G\'omez et
al. (2011): their RM map for January 1999 shows higher positive values on the
Southern side of the jet at the distance of components L and K (about 4~mas from
the core), but more negative values on the Southern side of the jet at the
distance of component O (about 2~mas from the core).

At first, this seems difficult to understand,
since the direction of an RM gradient associated with a helical {\bf B}
field is essentially determined by the direction of the rotation of the
central accretion disc and the direction of the poloidal field it winds up,
both of which we would expect to be constant in time.
We can offer several possible explanations 
for this result. We briefly discuss these below, 
and explain our reasoning for identifying the one that we think is the 
most likely (see also Mahmud et al. 2009). 

{\bf Torsional Oscillations of the Jet.}
One possible interpretation of oppositely directed core and jet transverse 
RM gradients, could be that the direction of the azimuthal {\bf B} field 
component changed as a result of torsional oscillations of jet 
(Bisnovatyi--Kogan 2007). Such torsional oscillations, which may help 
stabilize the jets, could cause a flip of the azimuthal {\bf B} field from 
time to time, or equivalently with distance from the core, given the jet 
outflow. In this scenario, we expect that the direction of the observed 
transverse RM gradients may reverse from time to time when the 
direction of the torsional oscillation reverses; this reversal pattern 
would presumably then propagate outward with the jet.  

{\bf Reversal of the ``pole'' facing the Earth.}
Another possible interpretation could be that the ``pole'' of the black hole 
facing the Earth reversed. One way to retain a transverse RM gradient in a 
helical magnetic field model but reverse the direction of this gradient, is 
if the direction of rotation of the central black hole (i.e.~the direction in 
which the field threading the accretion disc is ``wound up'') remains 
constant, but the ``pole'' of the black hole facing the Earth changes from 
North to South, or vice versa. To our knowledge, it is currently not known 
whether such polarity reversals are possible for the central black hole of 
AGN, or on what time scale they could occur.

{\bf Nested-helix B-field structure.} A simpler and more likely explanation 
is a magnetic-tower-type model, with poloidal magnetic flux and poloidal 
current concentrated around the central axis (Lynden-Bell 1996; 
Nakamura et al. 2006). Fundamental physics dictates that the magnetic-field
lines must close; in this picture, the magnetic field forms meridional 
loops that are anchored in the inner and outer parts of the accretion disc, 
which become twisted due to the differential rotation of the disc. 
This should essentially give rise to an ``inner'' helical B field near the 
jet axis and an ``outer'' helical field somewhat further from the jet axis. 
These two regions of helical field will be associated with oppositely directed 
RM gradients, and the total observed RM gradient will be determined by which 
region of helical field dominates the observed RMs. Thus, the presence of a 
change in the direction of the observed transverse RM gradient between the 
core/innermost jet and jet regions well resolved from the core could represent 
a transition from dominance of the inner to dominance of the outer helical 
{\bf B} fields in the total observed RM. This seems to provide the
simplest explanation for the RM-gradient reversals we observe in these
two objects.

Typically, we would expect the 
direction of the RM gradients in the core and jet (i.e., the regions whose 
net RM is determined by the inner/outer helical fields) to remain constant 
in time, since they should be determined by the source geometry and viewing 
angle. Mahmud et al. (2009) discuss the possibility that this 
type of ``nested helical field'' structure could also occasionally give rise 
to  changes in the direction of the observed RM gradients with time within
a given source. 

\section{Conclusion}
The polarization rotation-measure images for the two BL~Lac objects presented 
here provide new evidence in support of helical magnetic fields  associated
with the jets of these AGN, most importantly, the presence of transverse 
rotation measure gradients across the jets of both objects. There is also
a dominance of transverse {\bf B} fields in the jets 
of 0716+714 and 1749+701, and signs of `spine-sheath' polarization structures 
or orthogonal polarization offset toward one side of the jet  
in both these sources, consistent with the possibility that these jets
carry a helical magnetic-field component:
%Although some authors have
%suggested that polarization structures with a `spine' of orthogonal {\bf B}
%field and a `sheath' of longitudinal {\bf B} field could come about due to
%the joint effects of shock compression and interaction with the surrounding
%medium (e.g. Attridge et al. 1999), 
this type of structure can also come about 
naturally in the case of a helical jet {\bf B} field (e.g. Lyutikov et al. 
2005; Pushkarev et al.  2005). We interpret the observed transverse RM
gradients as being due to the systematic variation of the toroidal component
of a helical {\bf B} field across the jet (Blandford 1993). We note in this
connection that the transverse RM gradients in both 0716+714 and 1749+701 have opposite
signs on either side of the jet, making it impossible to explain the gradients
as an effect of changing thermal-electron density alone (there must be a change
in the direction of the line-of-sight magnetic field).  
%In the case of
%1749+701, we observe transverse RM gradients on both parsec and decaparsec
%scales, providing additional evidence that these are associated with the overall
%``global'' structure of the jet magnetic field.

We have also detected tentative transverse RM gradients in the region of the 
observed VLBI core in both BL~Lac objects, which can be interpreted as 
being associated with helical {\bf B} fields in the innermost jets of these
sources. Further, we have found a striking new feature of the RM distributions
in these objects: a reversal in the direction of the 
transverse RM gradients. Similar reversals can be seen in the RM images for
3C~120 presented by G\'omez et al. (2011).
At first, this seems difficult to understand,
since the direction of the RM gradient associated with a helical {\bf B}
field is essentially determined by the direction of the rotation of the
central accretion disc and the direction of the poloidal field it winds up.
We suggest that the most likely explanation for these reversals is that we
are dealing with a `nested-helical-field' structure such as that present in
magnetic-tower models, in which poloidal field lines emerging from the inner 
accretion disc form meridional loops that close in the outer part of the disc, 
with both sides of the loops (which have oppositely directed poloidal field 
components) getting `wound up' by the disc rotation. 

Further observations and studies of the RM-gradient reversals observed in
these objects can potentially provide key information about how the 
geometry of the magnetic fields in these AGN jets evolve, and may  
provide information on the jet dynamics and jet collimation. We are currently
using a variety of multi-frequency polarization VLBA data to search for
additional candidates for AGN jets displaying 
RM gradients and RM-gradient reversals on both parsec and decaparsec scales.

\section{Acknowledgements}

The research for this publication was supported by a Research Frontiers
Programme grant from Science Foundation Ireland and the Irish Research Council
for Science Engineering and Technology (IRCSET). The National Radio Astronomy 
Observatory is operated by Associated Universities Inc. We thank R.~Zavala for
kindly providing the modified version of the AIPS `RM' task used in this work.
We are also grateful to the referee for his careful reading of the paper and
thoughtful, competent and useful comments.

\section{Appendix: Monte Carlo Simulations}

We constructed a model source with a transverse RM gradient present 
across its jet, and carried out Monte Carlo simulations based on this model
source. The model source is cylindrical, with a fall-off in intensity on
either side of the cylinder axis, and along the axis of the cylinder from a
specified point located near one end of the cylinder (see Fig.~7). The resulting
appearance of the model emission region is broadly speaking ``core--jet-like''.

Model visibility data were generated for each of the six frequencies 
listed in Section~2.1 (4.6--15.4~GHz), including the effect of the transverse 
RM gradient in the $Q$ and $U$ visibility data, and these model visibility 
data were sampled at precisely the $(u,v)$ points at which 0716+714 was 
observed at each of the frequencies.  Random thermal noise and the effect 
of uncertainties in the EVPA calibration 
by up to $3^{\circ}$ were added to the sampled model visibilities. The 
amount of thermal noise added was chosen to yield rms values in the simulated
images that were comparable to those in our actual observations. 

Stokes $I$, 
$Q$ and $U$ images were constructed from these visibilities in CASA, using
the same beam as was used in the observations of 0716+714 presented here
($1.28\times 1.06$~mas in $PA = -0.84^{\circ}$, where the dimensions given
correspond to the full width at half maximum of the beam along its major and
minor axes). The polarization of the model was chosen to yield a degree of
polarization in the lower half of the convolved model image (the ``core'' region)
of about 5\% and a degree of polarization in the upper half of the convolved 
model image  of about 10\% -- similar to the observed values for 0716+714.  
The $Q$ and $U$ images were then used to construct the 
corresponding polarization angle (PANG) images at each frequency, which were, 
in turn, used to construct RM images in the usual way. Finally, Monte Carlo 
RM maps were constructed, based on 200 independent realizations of the thermal 
noise and EVPA calibration uncertainty. In each case, the RM values were output
to the RM map only in pixels in with the RM uncertainty indicated by the fitting
was less than 80~rad/m$^2$; this value was chosen so that no spurious pixels
were written to the output RM maps for any of the 200 realizations of the RM
distribution. Finally, an average RM map was derived by averaging together
all 200 individual realizations of the RM distribution.

This procedure was carried out for a number of model sources, all with a 
length of 1~mas and with transverse widths of 0.50, 0.35, 0.20. 0.10 and 
0.05~mas. A recent observation of 0716+714 with the {\emph RadioAstron} space
antenna and the European VLBI Network has measured the size of a feature in
the 6.2-cm core region to be 0.07~mas (Kardashev et al. 2013), and our narrowest
jet was designed to have a width somewhat smaller than this.

We considered two types of monotonic transverse RM gradients: 
uni-directional along the entire source structure, and oriented in one direction 
in the ``core'' region and in the opposite direction in the ``jet'' region,
i.e., showing a reversal. These Monte Carlo simulations complement those
carried out by Hovatta et al. (2012), in which simulated RM maps were made
from model data that did not contain RM gradients, to determine  the frequency
of spurious transverse RM gradients appearing in the simulated RM maps.

Examples of the total intensity maps of the model sources used in the simulation 
are shown in Fig. 7 in this Appendix, and the results of the RM Monte Carlo simulations 
are shown in Figs.~8--15 in this Appendix (we do not show the results for the
jet width of 0.50~mas, since these are very similar for the 0.35-mas jet width). 
The panels in Figs.~8--15
show (i) the RM map obtained by putting data without added 
thermal noise through the imaging procedure (i.e., the intrinsic RM distribution, 
but subject to errors due to the CLEAN process and limited $uv$ coverage); 
(ii) two examples of the individual ``noisy'' RM maps obtained. Note that the
colour scales for the three maps in a corresponding set have been  individually
chosen to highlight the RM patterns present, and may differ somewhat in some cases. 

In all cases, the RM gradients that
were introduced into the simulated data are visible in the ``noisy'' RM maps
that were obtained, even when the intrinsic width of the jet is approximately
1/20 of the beam full-width at half-maximum (FWHM). This may seem surprising, but
it is clearly demonstrated by the simulated data. The magnitude of the RM gradient
is reduced by the convolution more and more as the size of the beam relative to
the intrinsic size of the jet width increases, but the RM gradients that were
initially introduced into the simulated data remain visible. In the case of 
jet widths much less than the beam FWHM, the appearance of individual realizations 
can sometimes be fairly strongly distorted by noise; however, in all cases, 
averaging together all the individual realizations confirms the presence of the
RM gradients in the simulated images.
 
These results essentially indicate that it may not be necessary to 
impose a restriction on the width spanned by an observed RM gradient, {\emph provided 
that the difference between the RM values observed at opposite ends of the gradient 
is at least $3\sigma$}. This is consistent with the results of Murphy \& Gabuzda (2012),
who investigated the effect of resolution on transverse RM profiles. It is also 
consistent with Fig.~30 of Hovatta et al. (2012),
which shows that the fraction of ``false positives'', i.e., spurious RM gradients, 
that were obtained in their Monte Carlo simulations did not exceed $\simeq 1\%$ when 
a $3\sigma$ criterion was imposed for the RM gradient, even when the observed width
of the RM gradient was less than 1.5 beamwidths.  It becomes important to place 
some restriction on the width spanned by the gradient if the difference between 
the RM values being compared is less than $3\sigma$, as was also shown clearly by the 
Monte Carlo simulations of Hovatta et al.  (2012).

\begin{figure}
 \begin{minipage}[t]{8.7cm}
 \begin{center}
 \includegraphics[width=8.7cm,clip]{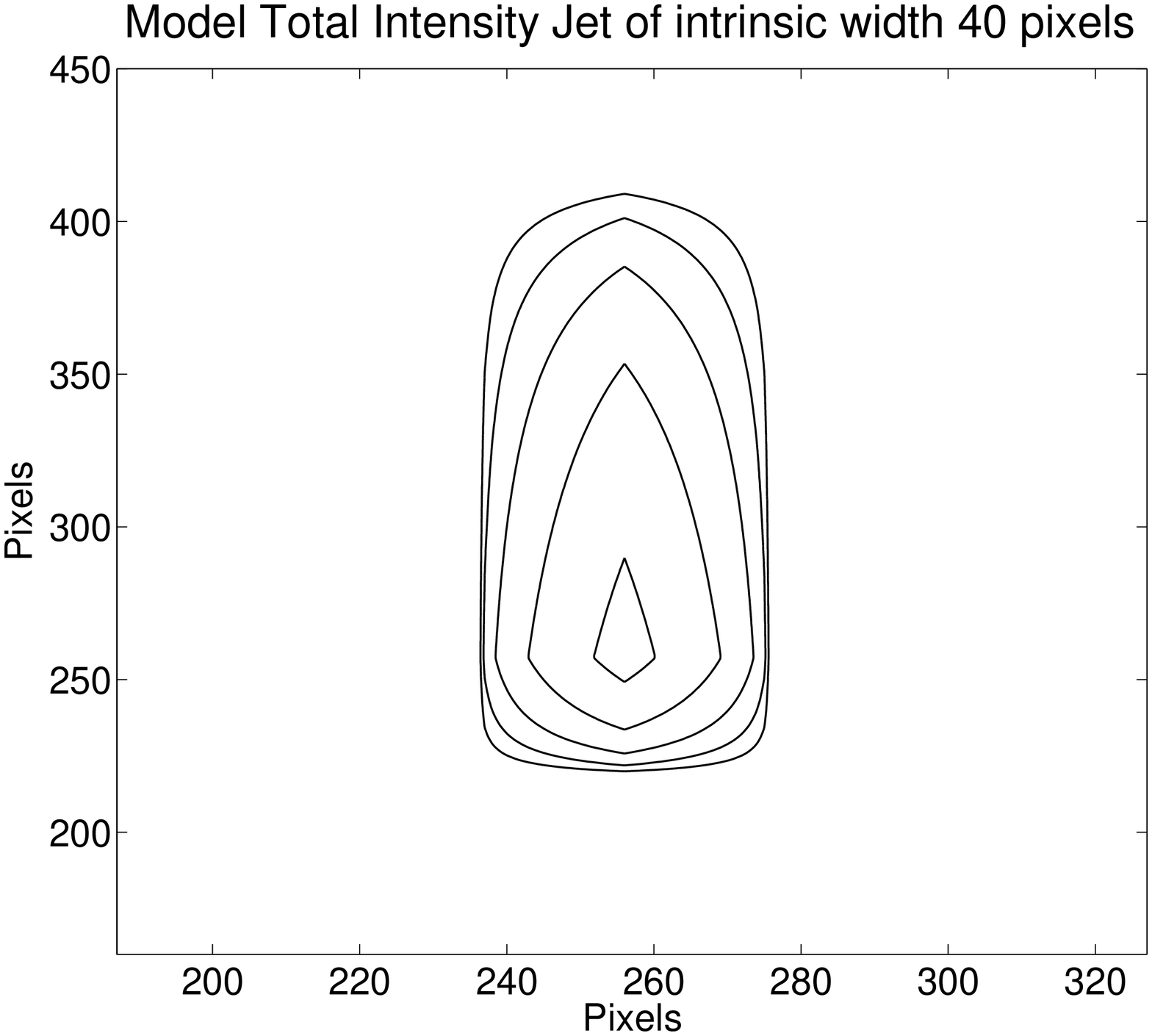}
 \end{center}
 \end{minipage}
 \begin{minipage}[t]{8.7cm}
 \begin{center}
 \includegraphics[width=8.7cm,clip]{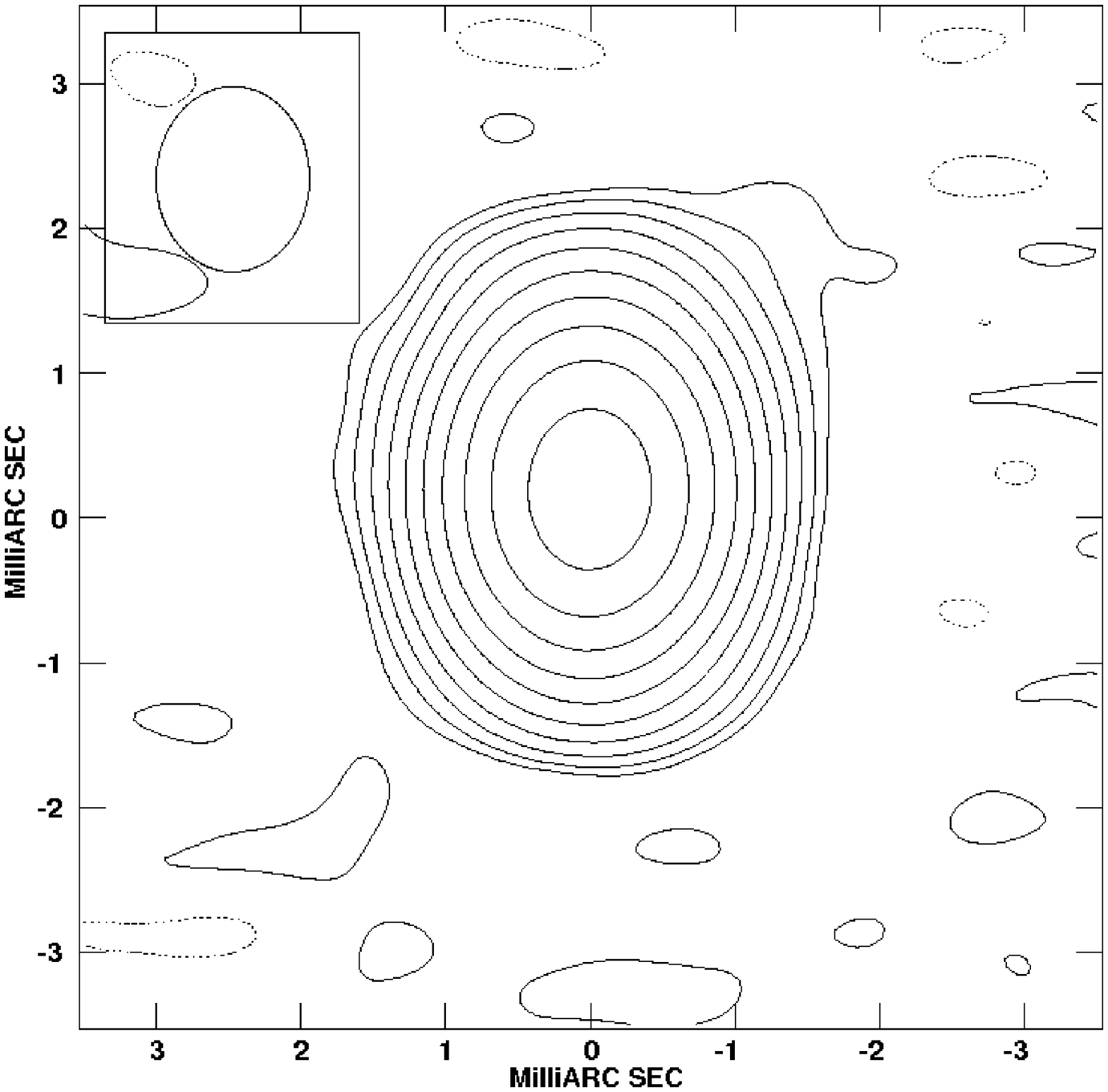}
 \end{center}
 \end{minipage}
\caption[Short caption for Figure 7]{\label{fig:MC_I} (Top) Intrinsic 
total intensity image of the model 
core--jet-like source with an intrinsic length of 1.0~mas (200~pixels) and an
intrinsic width of 0.20~mas (40~pixels), used for the Monte Carlo simulations.  
(Bottom) One realization of a ``noisy'' intensity map produced during the
simulations.
The convolving beam is 1.28~mas$\times$1.06~mas in PA = $-0.84^{\circ}$ 
(shown in the upp left-hand corner of the convolved image). The
peak of the unconvolved image is $5.62\times 10^{-4}$~Jy, and the
contours are 5, 10, 20, 40, and 80\% of the peak. The peak of the
convolved image is 1.11~Jy/beam, and the contours are $-0.125$, 0.125,
0.25, 0.5, 1, 2, 4, 8, 16, 32, and 64\% of the peak.  }
\end{figure}

%\begin{figure*}
% \begin{minipage}[t]{8.7cm}
% \begin{center}
% \includegraphics[width=8.7cm,clip]{MNR100_RMNN.eps}
% \end{center}
% \end{minipage}
% \begin{minipage}[t]{8.7cm}
% \begin{center}
% \includegraphics[width=8.7cm,clip]{MNR100_RMAVG.eps}
% \end{center}
% \end{minipage}
%\begin{minipage}[t]{8.7cm}
% \begin{center}
% \includegraphics[width=8.7cm,clip]{MNR100_RM90.eps}
% \end{center}
%\end{minipage}
%\begin{minipage}[t]{8.7cm}
% \begin{center}
% \includegraphics[width=8.7cm,clip]{MNR100_RM163.eps}
% \end{center}
%\end{minipage}
%\caption[Short caption for Figure 8]{\label{fig:MNR100} Results of
%Monte Carlo simulations using model core--jet sources with uniformly
%directed transverse
%RM gradients. The intrinsic width of the jet (RM gradient) is 0.50~mas.
%The convolving beam (1.28~mas$\times$1.06~mas in PA = $-0.84^{\circ}$)
%is shown in the lower left-hand corner of each panel. The top
%left panel shows the RM image obtained by processing the model
%data as usual, but without adding random noise  or EVPA calibration
%uncertainty; pixels with RM uncertainties exceeding 10~rad/m$^2$ were
%blanked.  The bottom two panels show two examples of the 200 individual 
%RM images obtained during the simulations; pixels with RM uncertainties
%exceeding 80~rad/m$^2$ were blanked.  The top right panel shows the average 
%of the 200 individual Monte Carlo RM images obtained, each with noise and 
%EVPA calibration uncertainty added.  }
%\end{figure*}

\begin{figure}
% \begin{minipage}[t]{5.5cm}
 \begin{center}
 \includegraphics[width=9.2cm,clip]{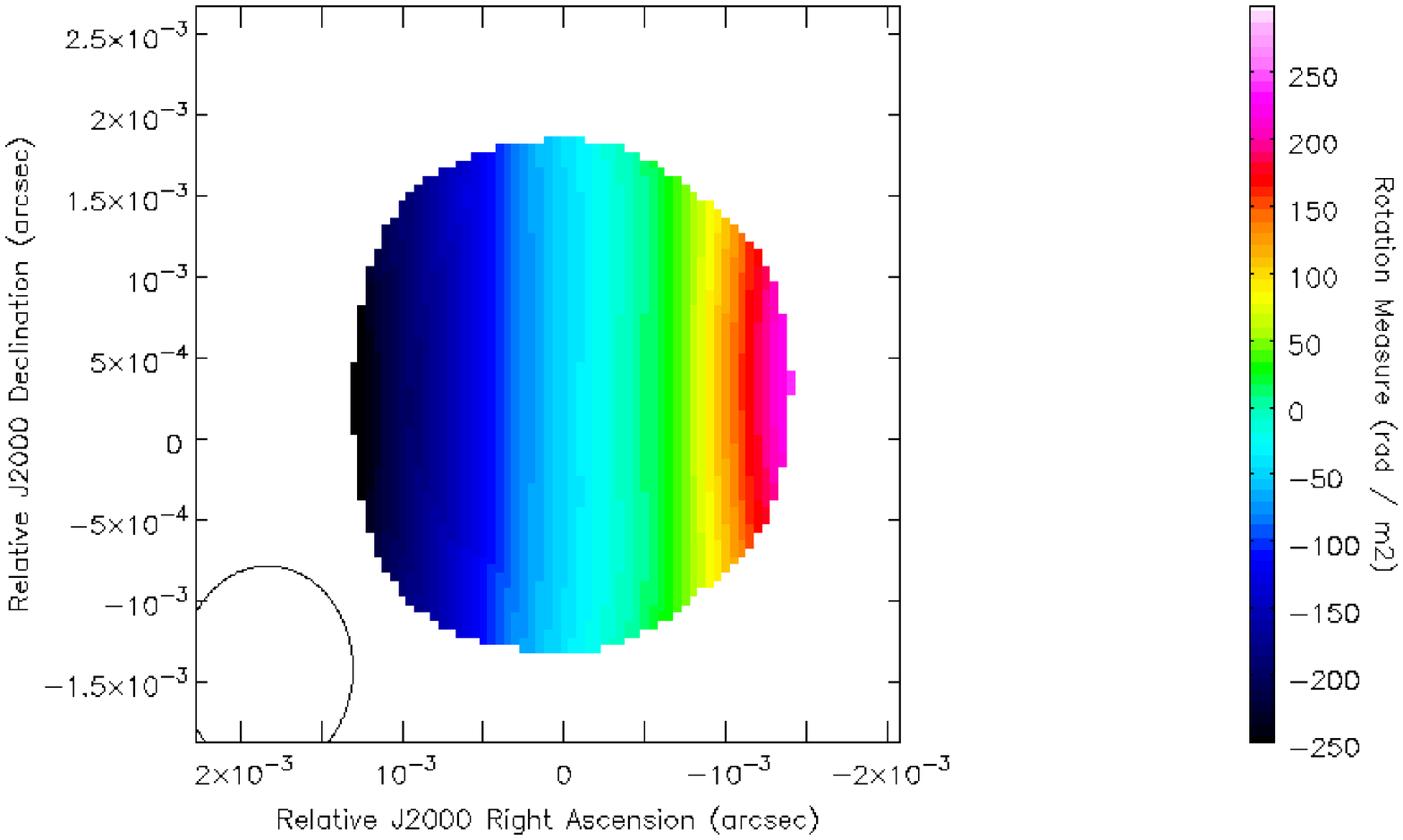}
 \end{center}
% \end{minipage}
% \begin{minipage}[t]{8.7cm}
% \begin{center}
% \includegraphics[width=8.7cm,clip]{MNR66_RMAVG.eps}
% \end{center}
% \end{minipage}
%\begin{minipage}[t]{5.5cm}
 \begin{center}
 \includegraphics[width=9.2cm,clip]{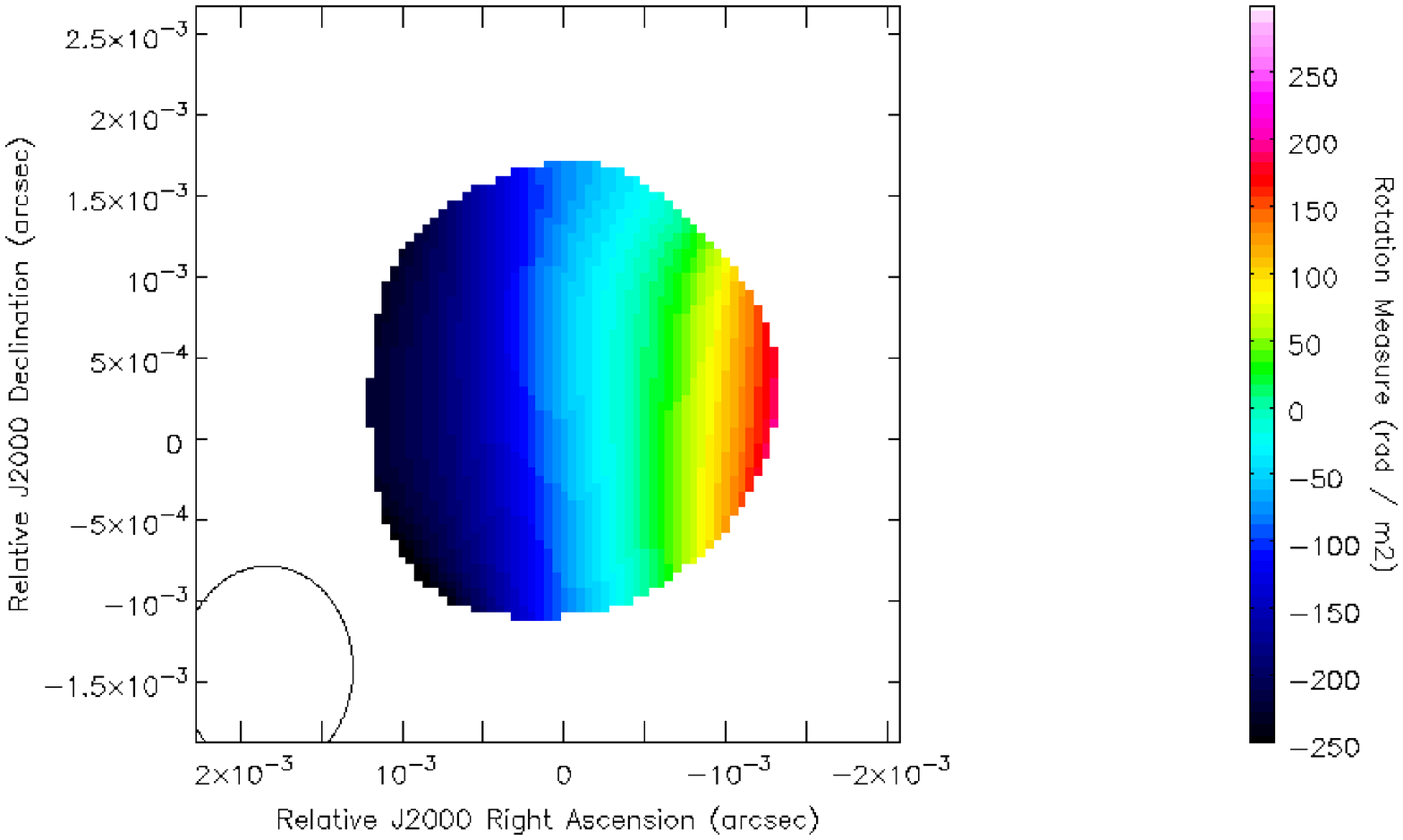}
 \end{center}
%\end{minipage}
%\begin{minipage}[t]{5.5cm}
 \begin{center}
 \includegraphics[width=9.2cm,clip]{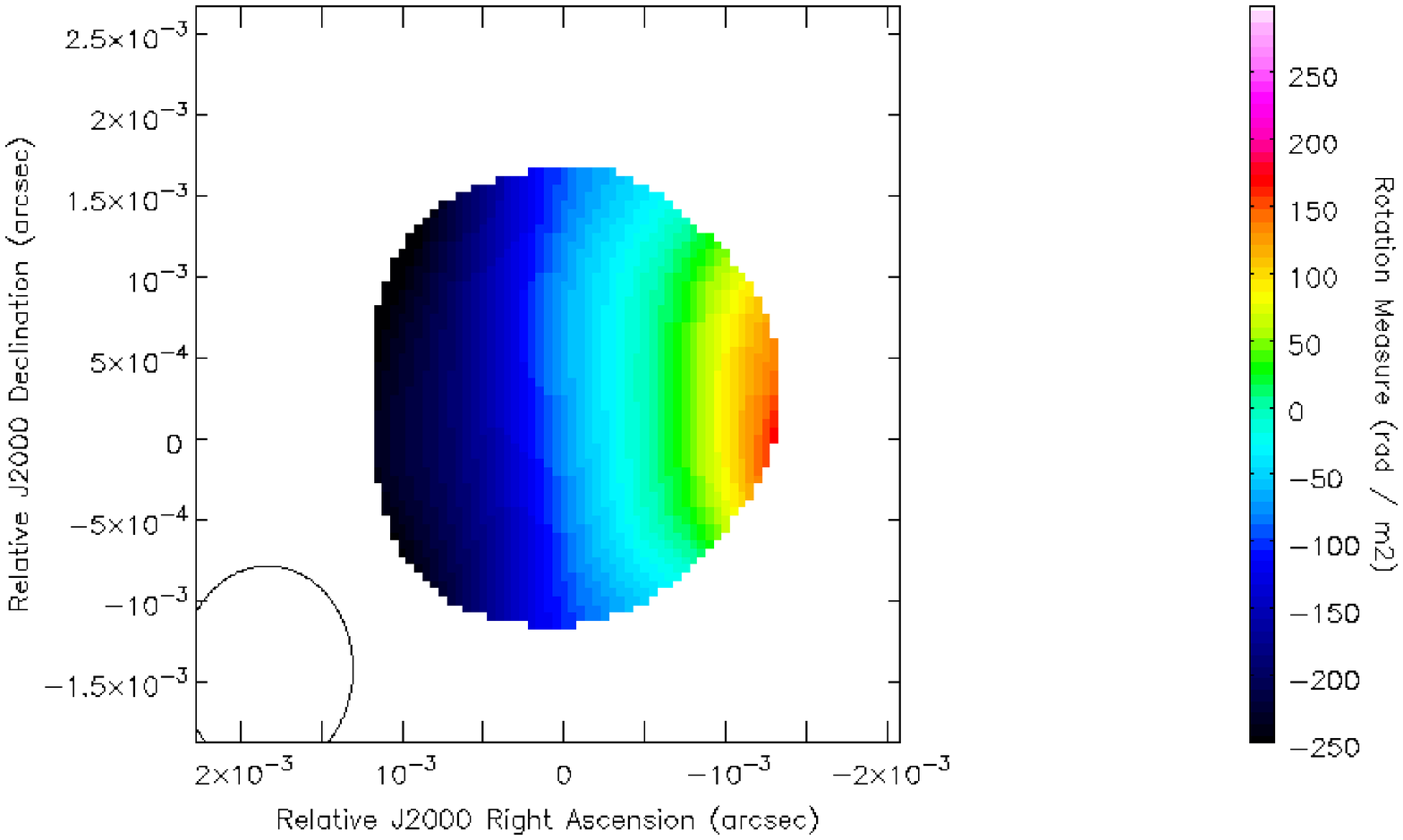}
 \end{center}
%\end{minipage}
\caption[Short caption for Figure 8]{\label{fig:MNR66} Results of
Monte Carlo simulations using model core--jet sources with uniformly
directed transverse
RM gradients. The intrinsic width of the jet (RM gradient) is 0.35~mas.
The convolving beam (1.28~mas$\times$1.06~mas in PA = $-0.84^{\circ}$)
is shown in the lower left-hand corner of each panel. The top
panel shows the RM image obtained by processing the model
data as usual, but without adding random noise  or EVPA calibration
uncertainty; pixels with RM uncertainties exceeding 10~rad/m$^2$ were
blanked.  The remaining two panels show two examples of the 200 individual 
RM images obtained during the simulations; pixels with RM uncertainties
exceeding 80~rad/m$^2$ were blanked. }
%\caption[Short caption for Figure 9]{\label{fig:MNR66} Same as Fig.~8
%for a core--jet source with the same length but an intrinsic jet width
%of 0.35~mas.}
\end{figure}

\begin{figure}
% \begin{minipage}[t]{8.7cm}
 \begin{center}
 \includegraphics[width=9.2cm,clip]{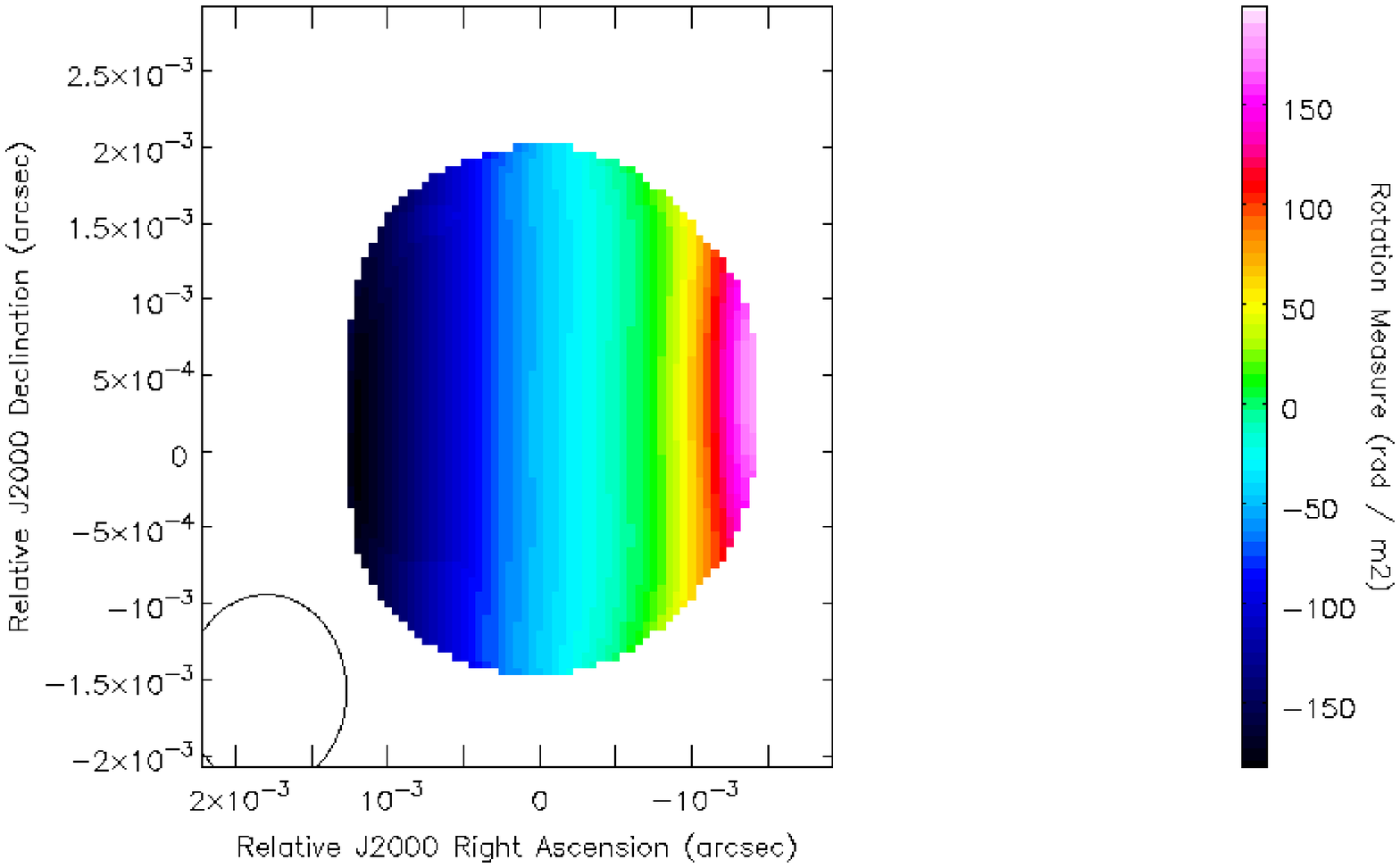}
 \end{center}
% \end{minipage}
% \begin{minipage}[t]{8.7cm}
% \begin{center}
% \includegraphics[width=8.7cm,clip]{MNR40_RMAVG.eps}
% \end{center}
% \end{minipage}
%\begin{minipage}[t]{8.7cm}
 \begin{center}
 \includegraphics[width=9.2cm,clip]{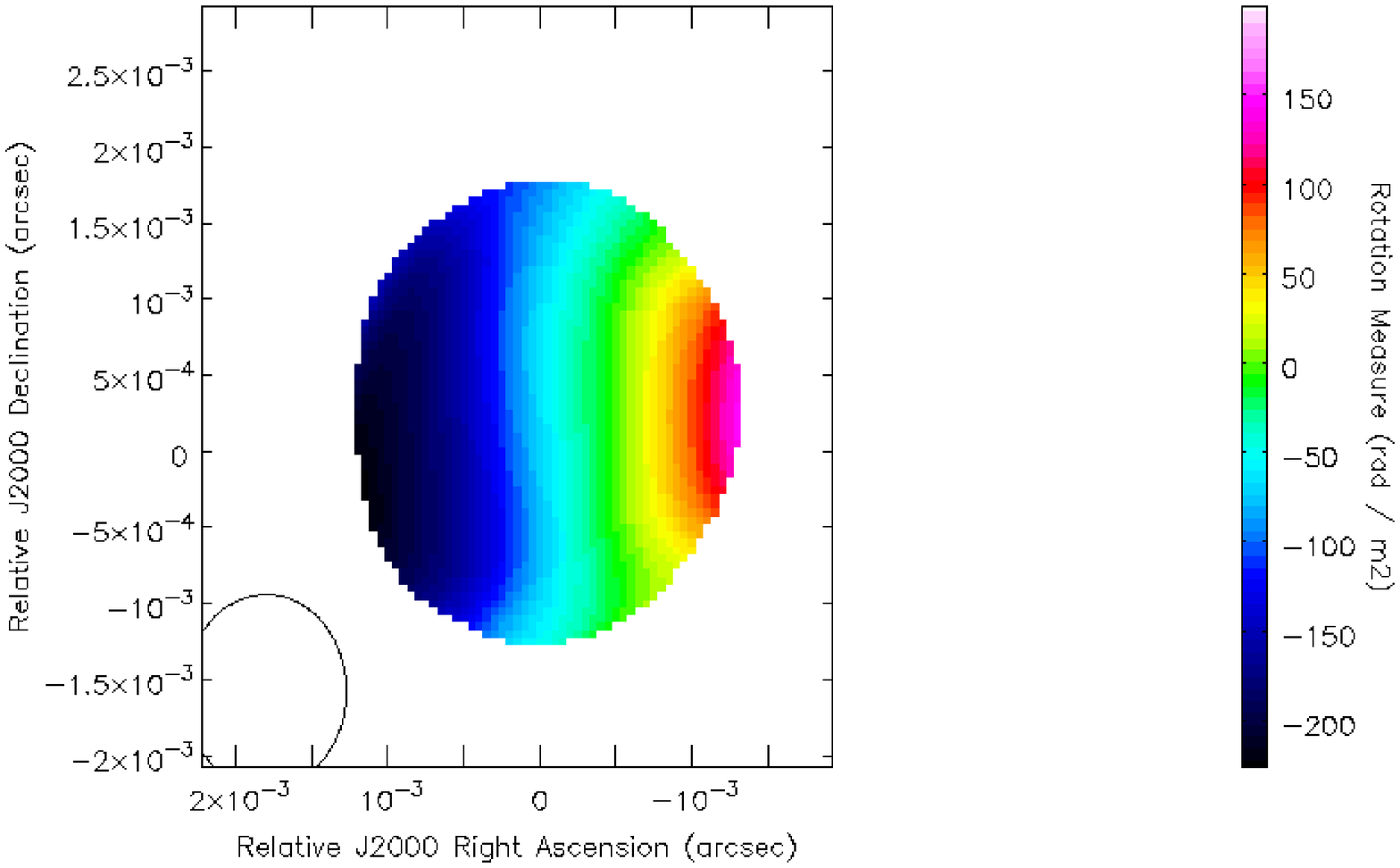}
 \end{center}
%\end{minipage}
%\begin{minipage}[t]{8.7cm}
 \begin{center}
 \includegraphics[width=9.2cm,clip]{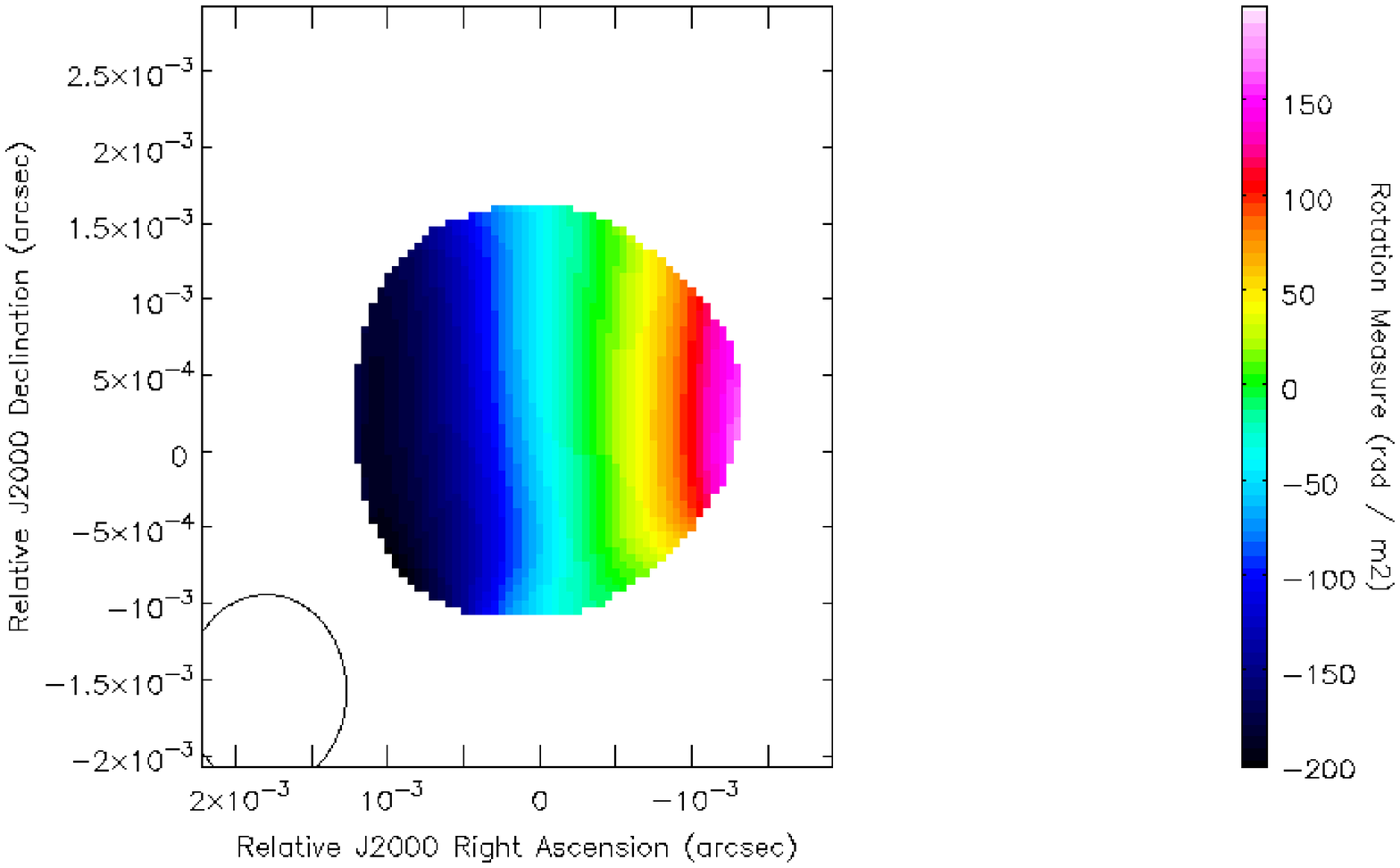}
 \end{center}
%\end{minipage}
\caption[Short caption for Figure 9]{\label{fig:MNR40} Same as Fig.~8
for a core--jet source with the same length but an intrinsic jet width
of 0.20~mas.  }
\end{figure}

\begin{figure}
% \begin{minipage}[t]{8.7cm}
 \begin{center}
 \includegraphics[width=9.2cm,clip]{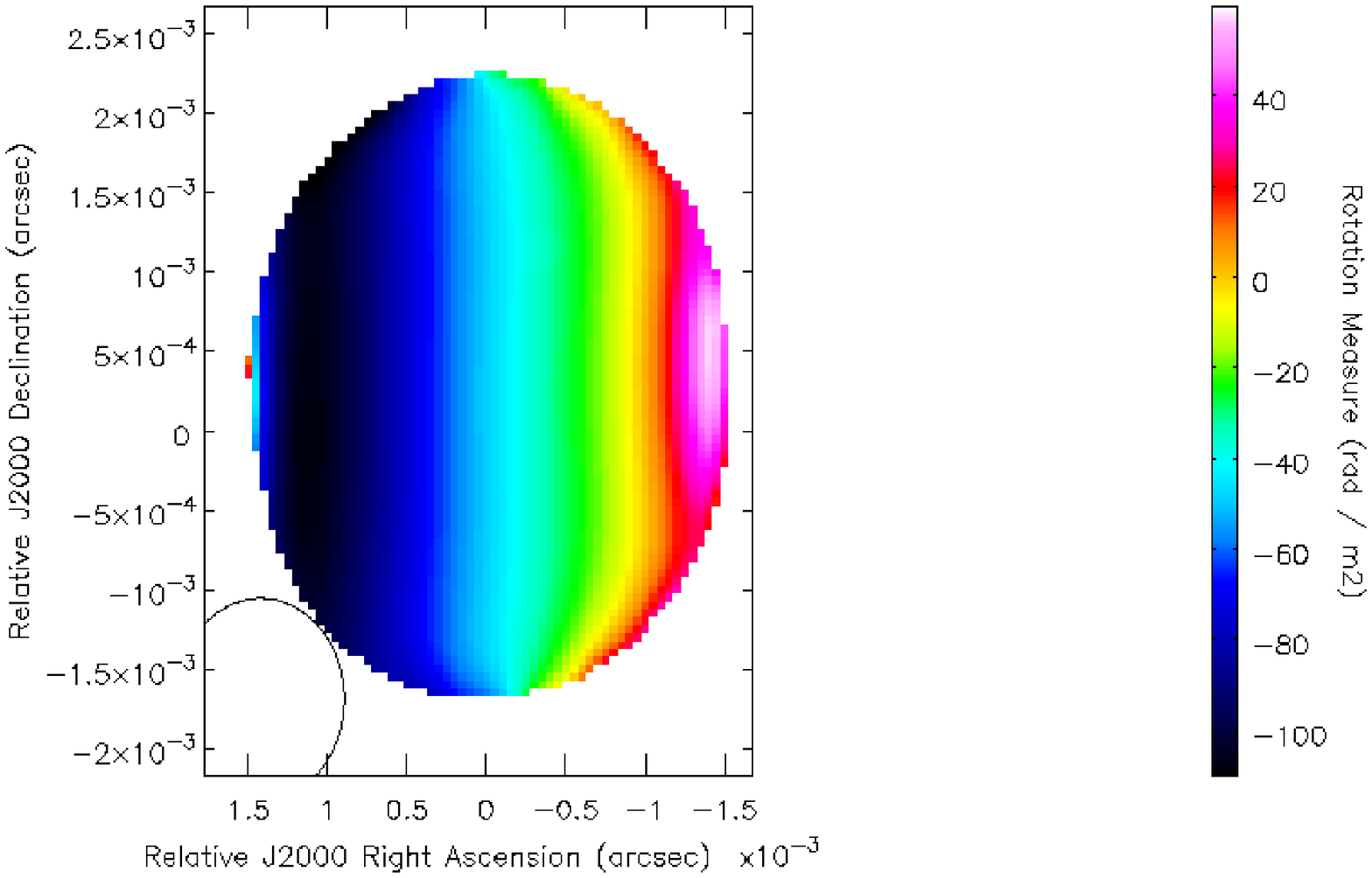}
 \end{center}
% \end{minipage}
% \begin{minipage}[t]{8.7cm}
% \begin{center}
% \includegraphics[width=8.7cm,clip]{MNR20_RMAVG.eps}
% \end{center}
% \end{minipage}
%\begin{minipage}[t]{8.7cm}
 \begin{center}
 \includegraphics[width=9.2cm,clip]{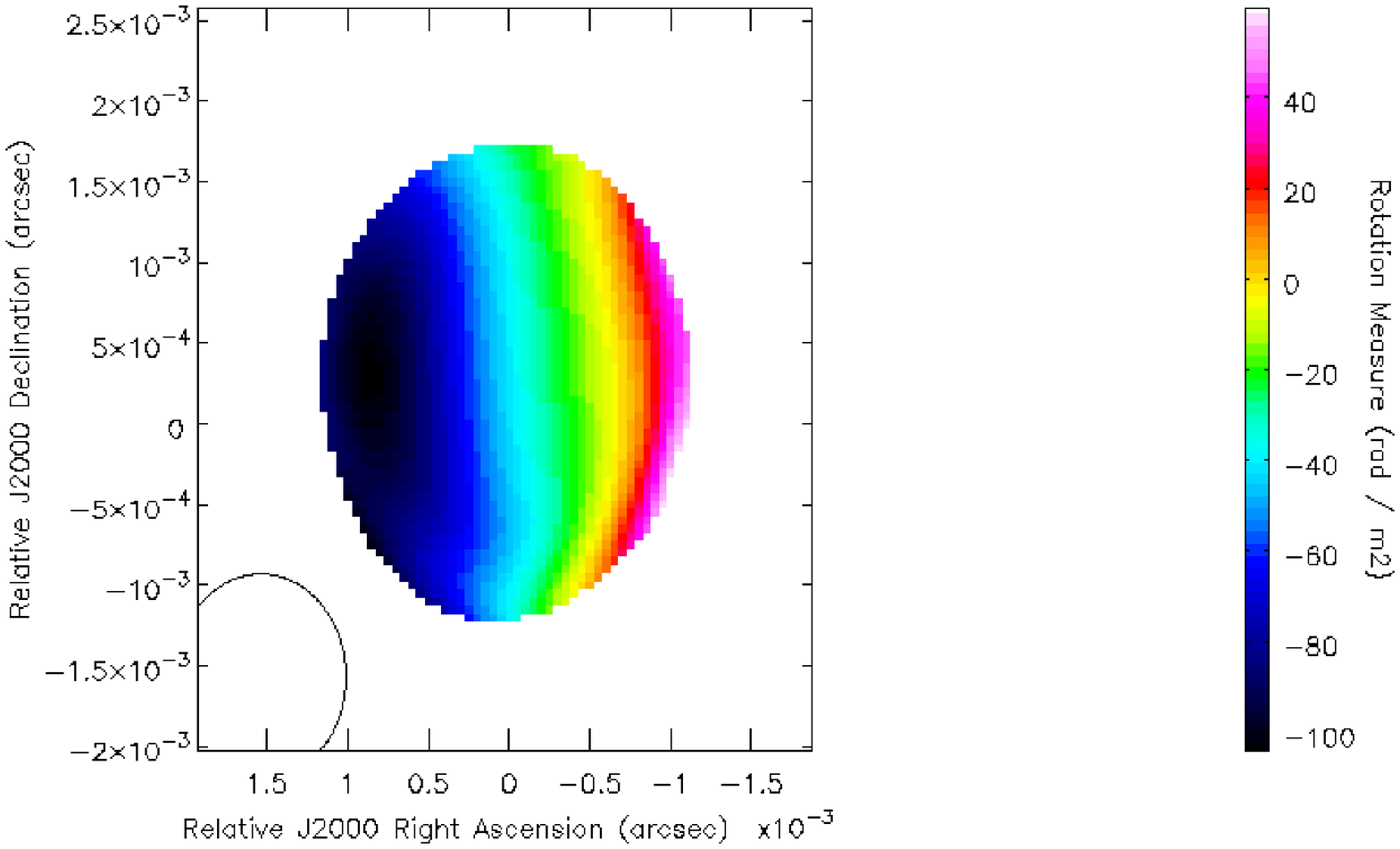}
 \end{center}
%\end{minipage}
%\begin{minipage}[t]{8.7cm}
 \begin{center}
 \includegraphics[width=9.2cm,clip]{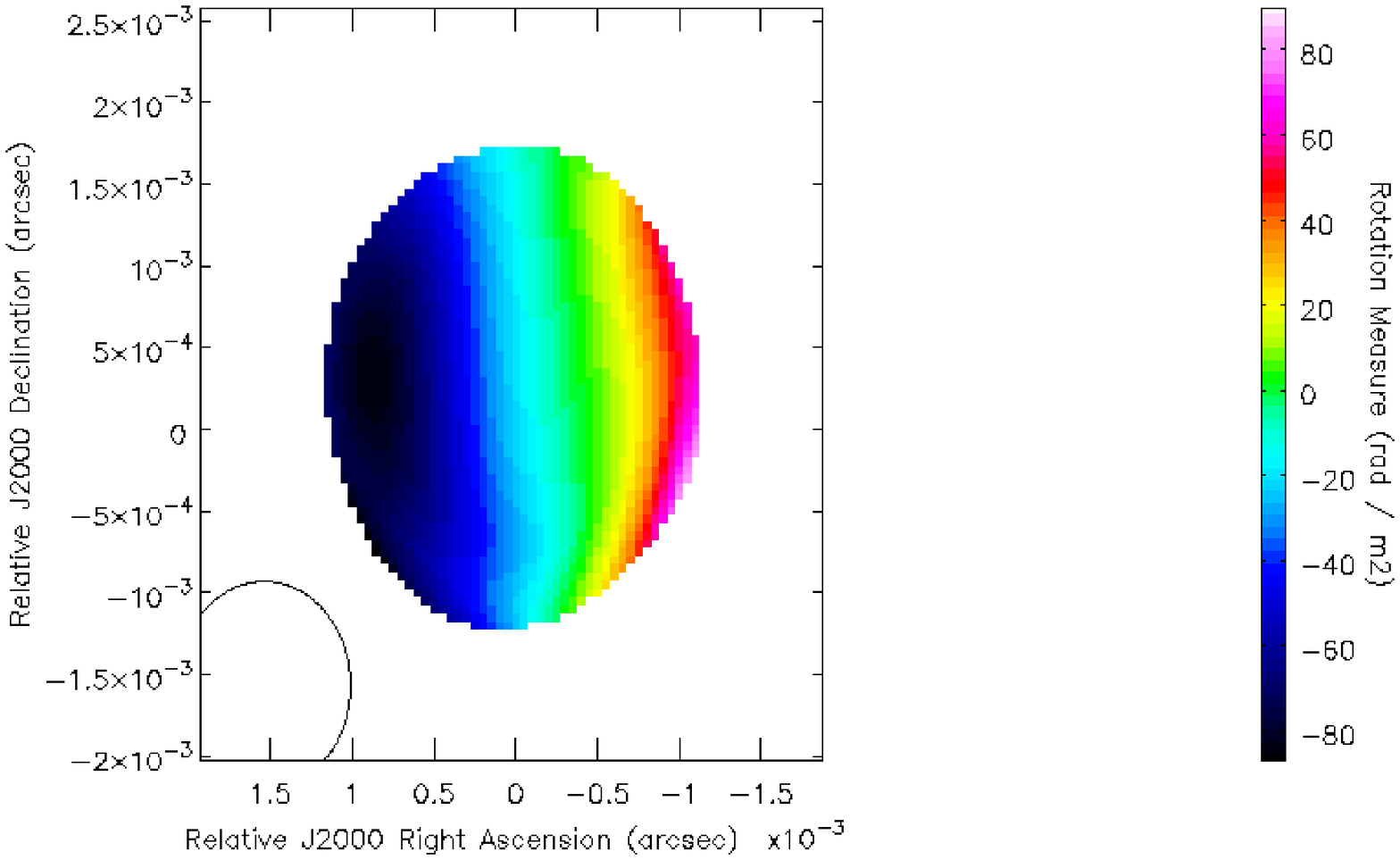}
 \end{center}
%\end{minipage}
\caption[Short caption for Figure 10]{\label{fig:MNR20} Same as Fig.~8
for a core--jet source with the same length but an intrinsic jet width
of 0.10~mas.  }
\end{figure}

\begin{figure}
% \begin{minipage}[t]{8.7cm}
 \begin{center}
 \includegraphics[width=9.2cm,clip]{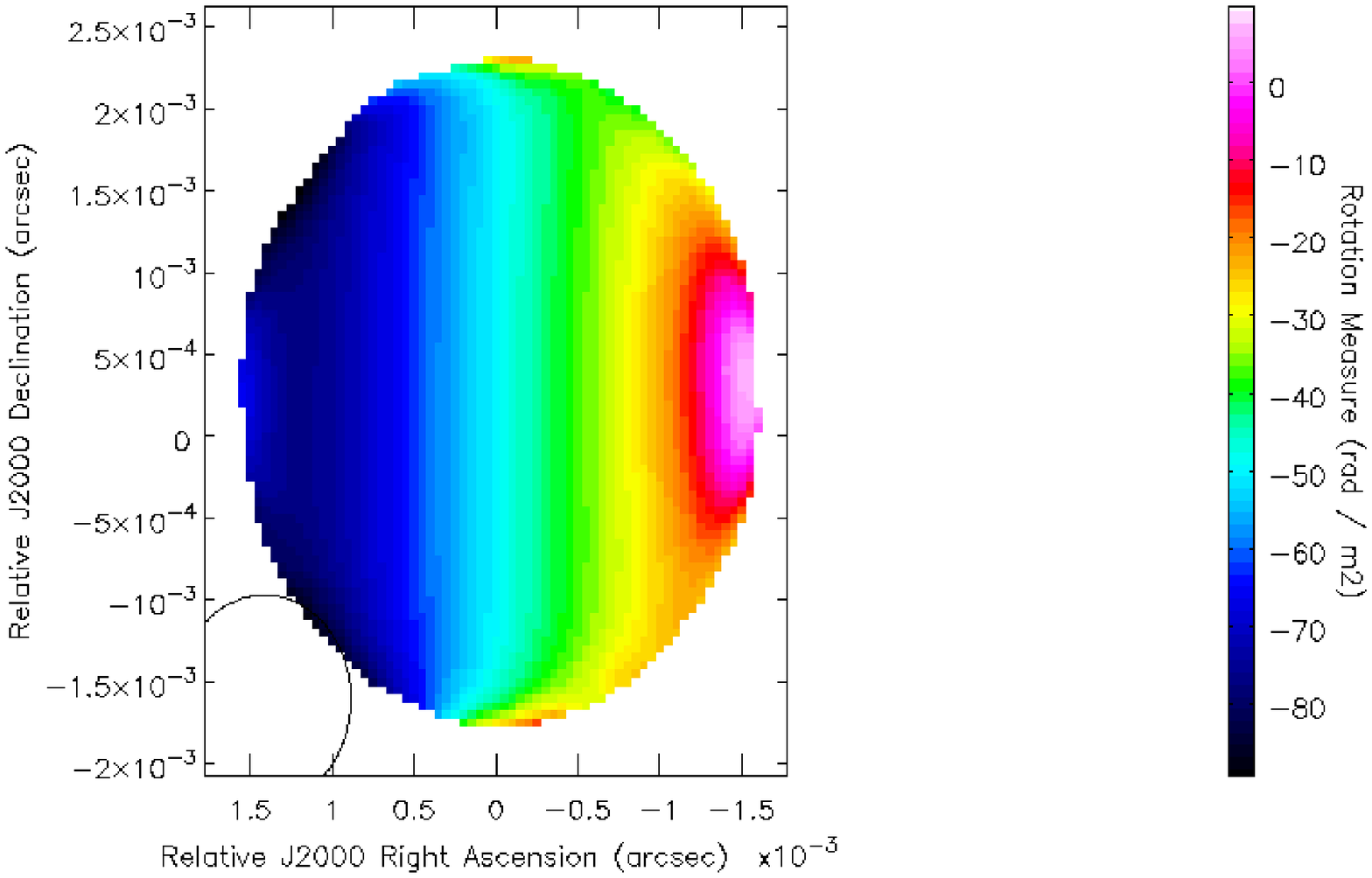}
 \end{center}
% \end{minipage}
% \begin{minipage}[t]{8.7cm}
% \begin{center}
% \includegraphics[width=8.7cm,clip]{MNR10_RMAVG.eps}
% \end{center}
% \end{minipage}
%\begin{minipage}[t]{8.7cm}
 \begin{center}
 \includegraphics[width=9.2cm,clip]{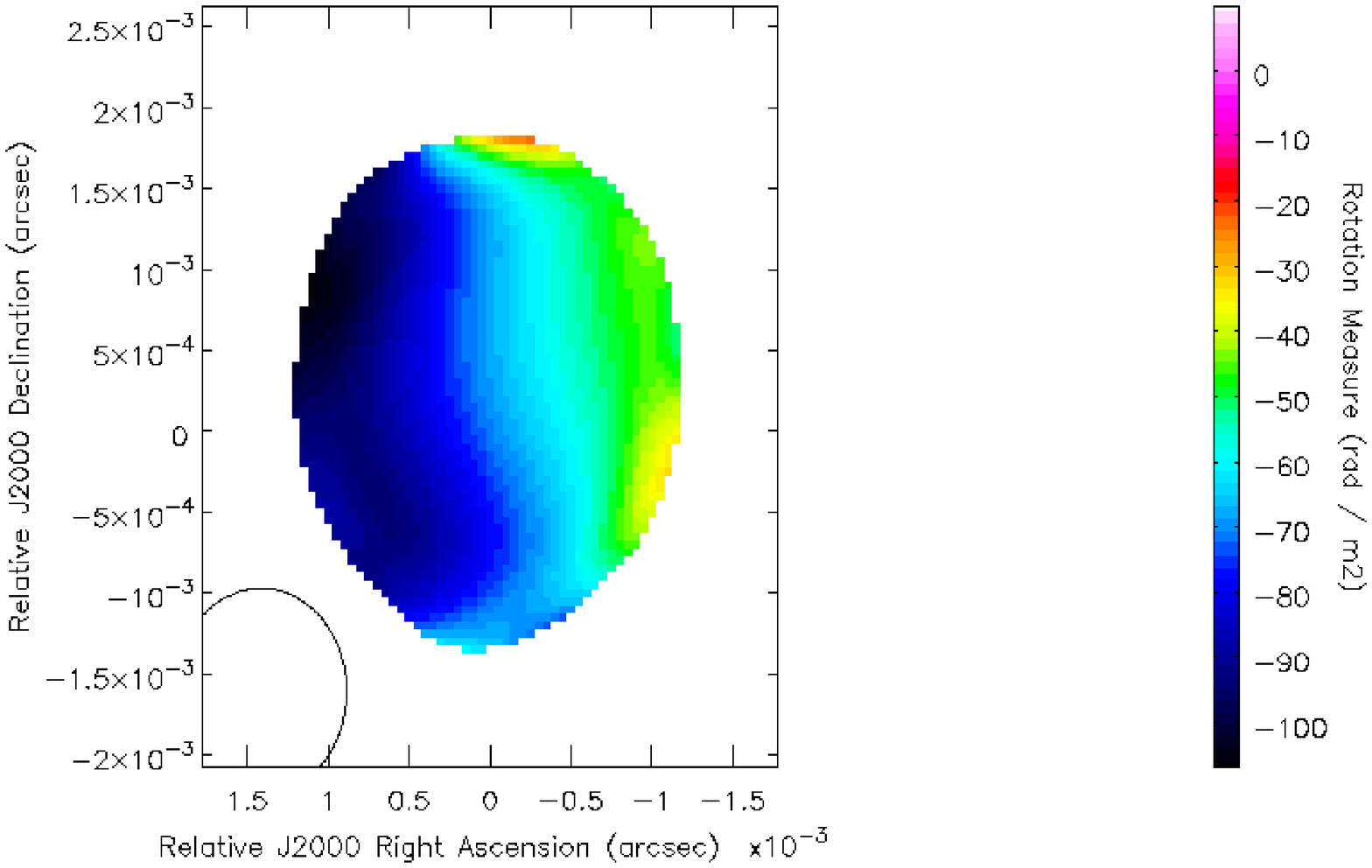}
 \end{center}
%\end{minipage}
%\begin{minipage}[t]{8.7cm}
 \begin{center}
 \includegraphics[width=9.2cm,clip]{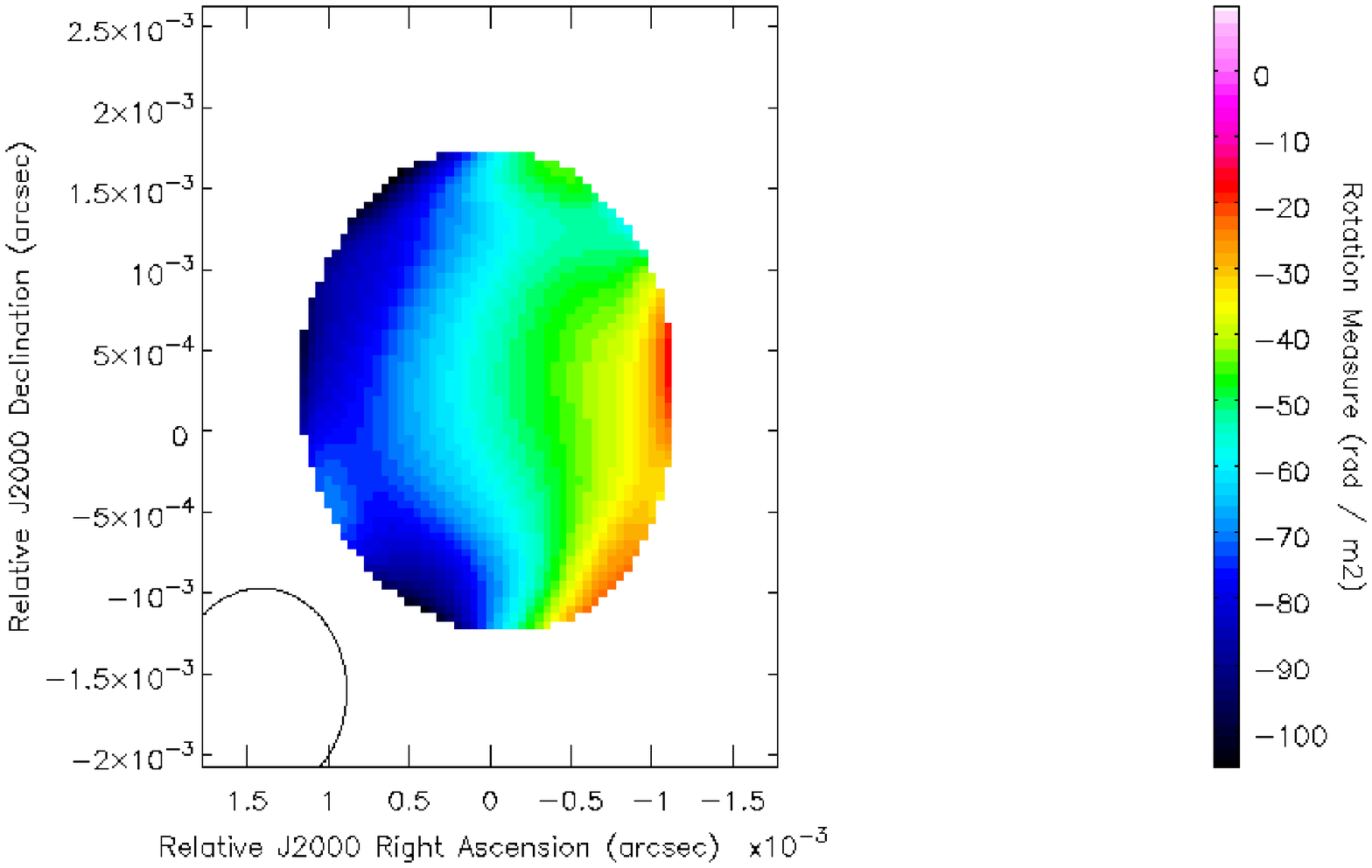}
 \end{center}
%\end{minipage}
\caption[Short caption for Figure 11]{\label{fig:MNR10} Same as Fig.~8
for a core--jet source with the same length but an intrinsic jet width
of 0.05~mas.  }
\end{figure}

%\begin{figure}
% \begin{minipage}[t]{8.7cm}
% \begin{center}
% \includegraphics[width=8.7cm,clip]{MRCG100NN_RM.eps}
% \end{center}
% \end{minipage}
% \begin{minipage}[t]{8.7cm}
% \begin{center}
% \includegraphics[width=8.7cm,clip]{MRCG100_RMAVG.eps}
% \end{center}
% \end{minipage}
%\begin{minipage}[t]{8.7cm}
% \begin{center}
% \includegraphics[width=8.7cm,clip]{MRCG100_RM113.eps}
% \end{center}
%\end{minipage}
%\begin{minipage}[t]{8.7cm}
% \begin{center}
% \includegraphics[width=8.7cm,clip]{MRCG100_RM137.eps}
% \end{center}
%\end{minipage}
%\caption[Short caption for Figure 13]{\label{fig:MRCG100} Results of
%Monte Carlo simulations using model core--jet sources with oppositely
%directed transverse RM gradients in the core region and inner jet. 
%The intrinsic width of the jet (RM gradient) is 0.50~mas.
%The convolving beam (1.28~mas$\times$1.06~mas in PA = $-0.84^{\circ}$)
%is shown in the lower left-hand corner of each panel. The top
%left panel shows the RM image obtained by processing the model
%data as usual, but without adding random noise  or EVPA calibration
%uncertainty; pixels with RM uncertainties exceeding 10~rad/m$^2$ were
%blanked.  The bottom two panels show two examples of the 200 individual 
%RM images obtained during the simulations; pixels with RM uncertainties
%exceeding 80~rad/m$^2$ were blanked.  The top right panel shows the average 
%of the 200 individual Monte Carlo RM images obtained, each with noise and 
%EVPA calibration uncertainty added.  }
%\end{figure}

\begin{figure}
% \begin{minipage}[t]{8.7cm}
 \begin{center}
 \includegraphics[width=9.2cm,clip]{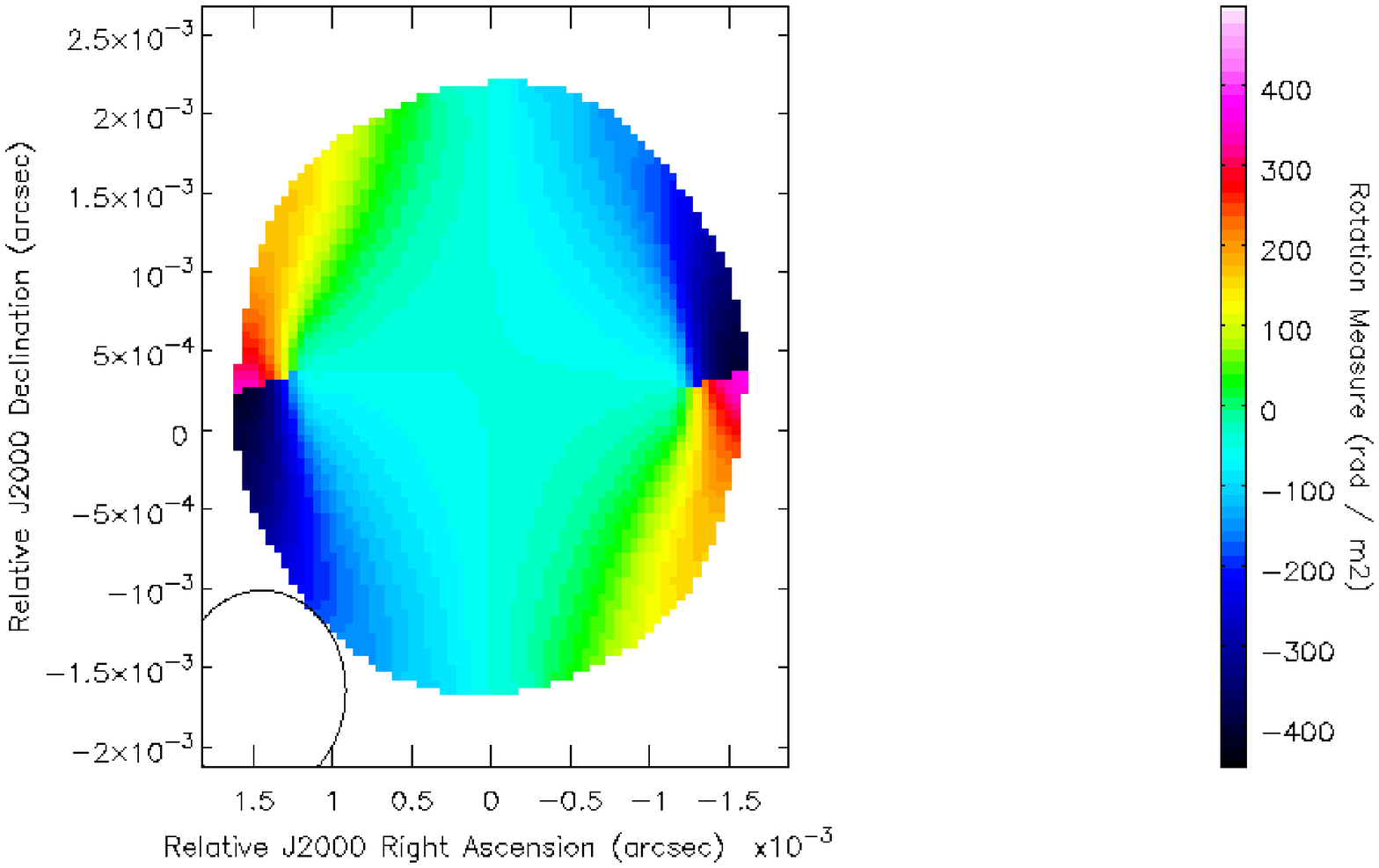}
 \end{center}
% \end{minipage}
% \begin{minipage}[t]{8.7cm}
% \begin{center}
% \includegraphics[width=8.7cm,clip]{MRCG66_RMAVG.eps}
% \end{center}
% \end{minipage}
%\begin{minipage}[t]{8.7cm}
 \begin{center}
 \includegraphics[width=9.2cm,clip]{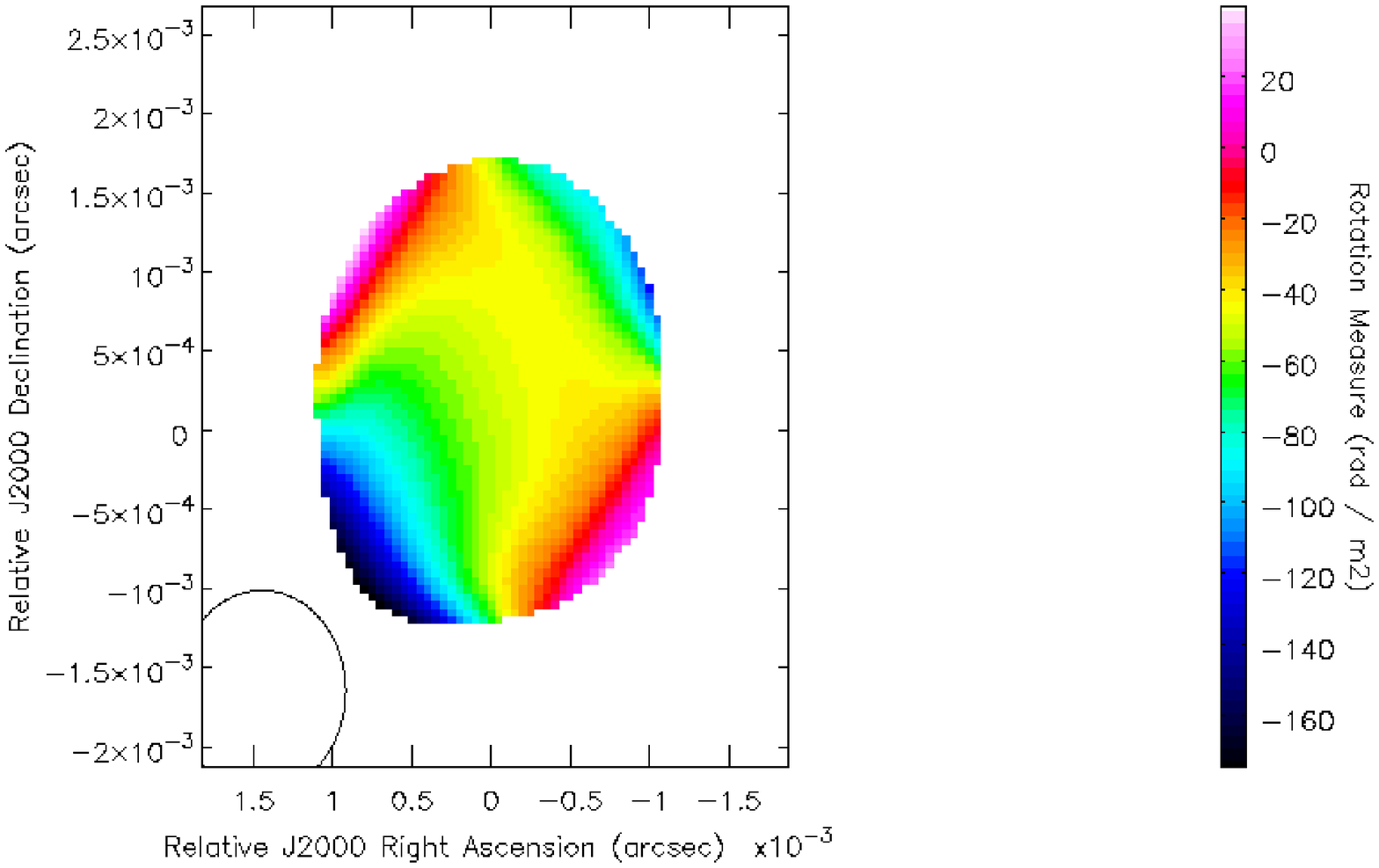}
 \end{center}
%\end{minipage}
%\begin{minipage}[t]{8.7cm}
 \begin{center}
 \includegraphics[width=9.2cm,clip]{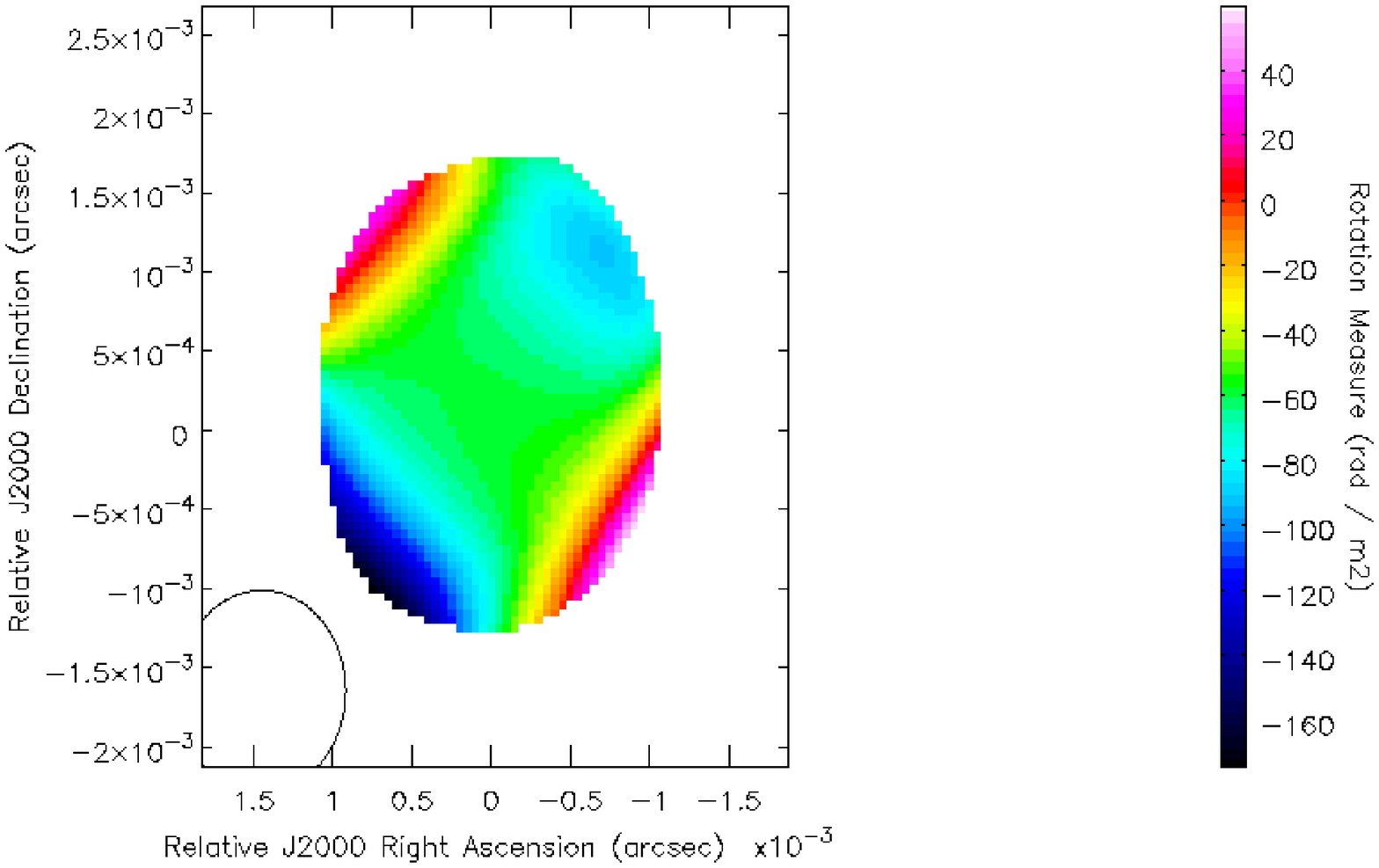}
 \end{center}
%\end{minipage}
\caption[Short caption for Figure 12]{\label{fig:MRCG100} Results of
Monte Carlo simulations using model core--jet sources with oppositely
directed transverse RM gradients in the core region and inner jet. 
The intrinsic width of the jet (RM gradient) is 0.35~mas.
The convolving beam (1.28~mas$\times$1.06~mas in PA = $-0.84^{\circ}$)
is shown in the lower left-hand corner of each panel. The top
panel shows the RM image obtained by processing the model
data as usual, but without adding random noise  or EVPA calibration
uncertainty; pixels with RM uncertainties exceeding 10~rad/m$^2$ were
blanked.  The remaining two panels show two examples of the 200 individual 
RM images obtained during the simulations; pixels with RM uncertainties
exceeding 80~rad/m$^2$ were blanked.  
%The top right panel shows the average 
%of the 200 individual Monte Carlo RM images obtained, each with noise and 
%EVPA calibration uncertainty added.  
}
%\caption[Short caption for Figure 13]{\label{fig:MCG66} Same as Fig.~13
%for a core--jet source with the same length but an intrinsic jet width
%of 0.35~mas.  }
\end{figure}

\begin{figure}
% \begin{minipage}[t]{8.7cm}
 \begin{center}
 \includegraphics[width=9.2cm,clip]{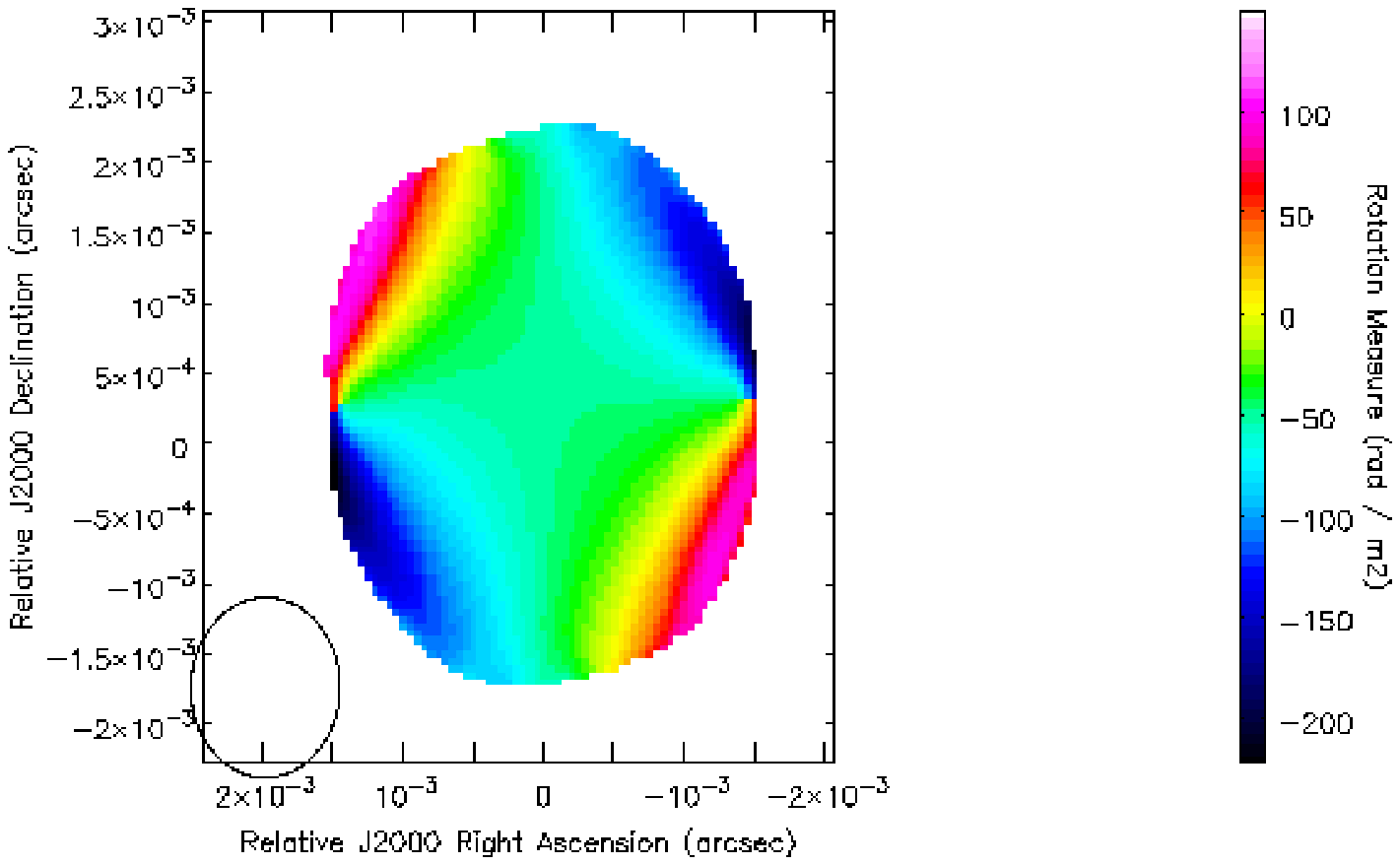}
 \end{center}
% \end{minipage}
% \begin{minipage}[t]{8.7cm}
% \begin{center}
% \includegraphics[width=8.7cm,clip]{MRCG40_RMAVG.eps}
% \end{center}
% \end{minipage}
%\begin{minipage}[t]{8.7cm}
 \begin{center}
 \includegraphics[width=9.2cm,clip]{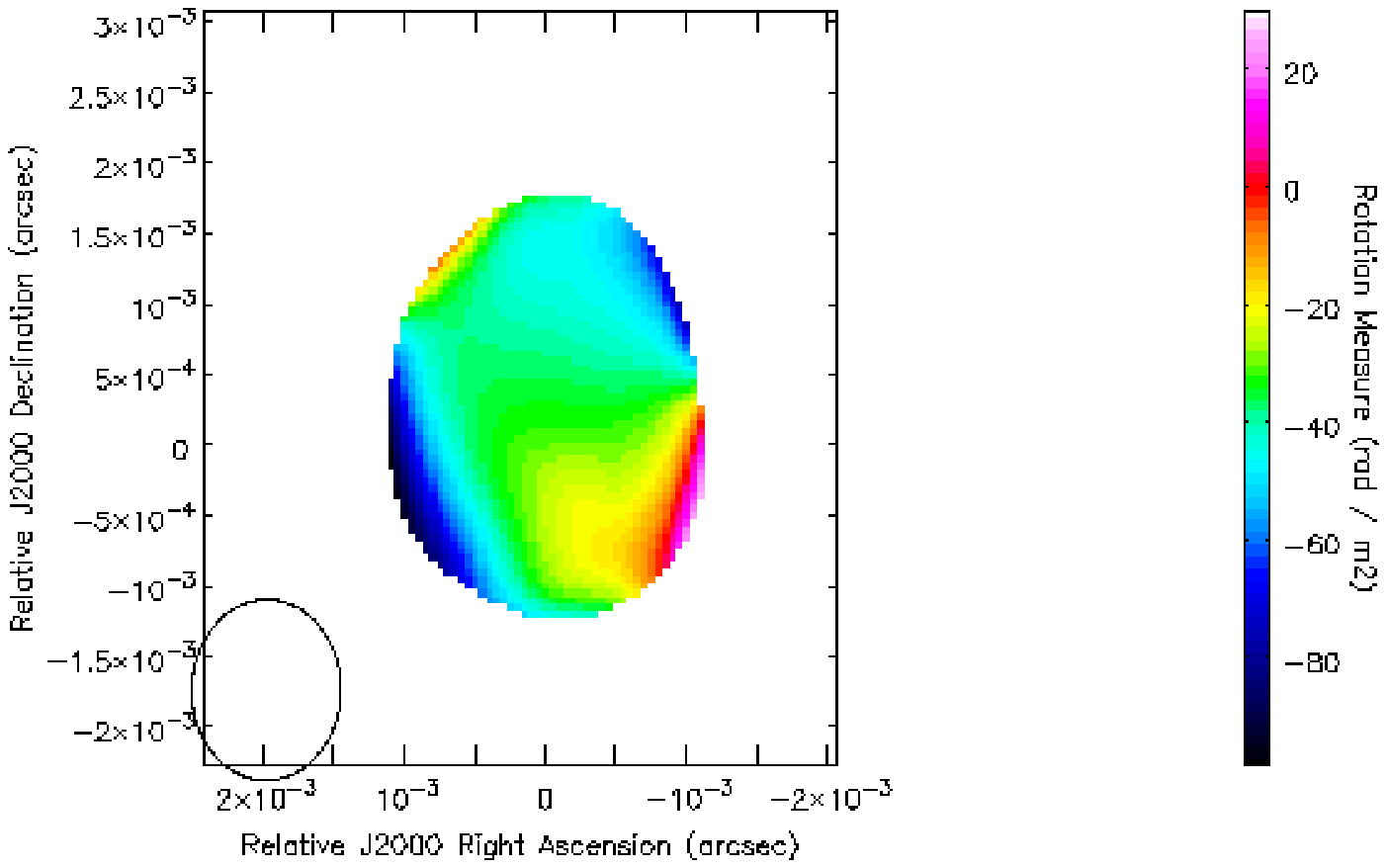}
 \end{center}
%\end{minipage}
%\begin{minipage}[t]{8.7cm}
 \begin{center}
 \includegraphics[width=9.2cm,clip]{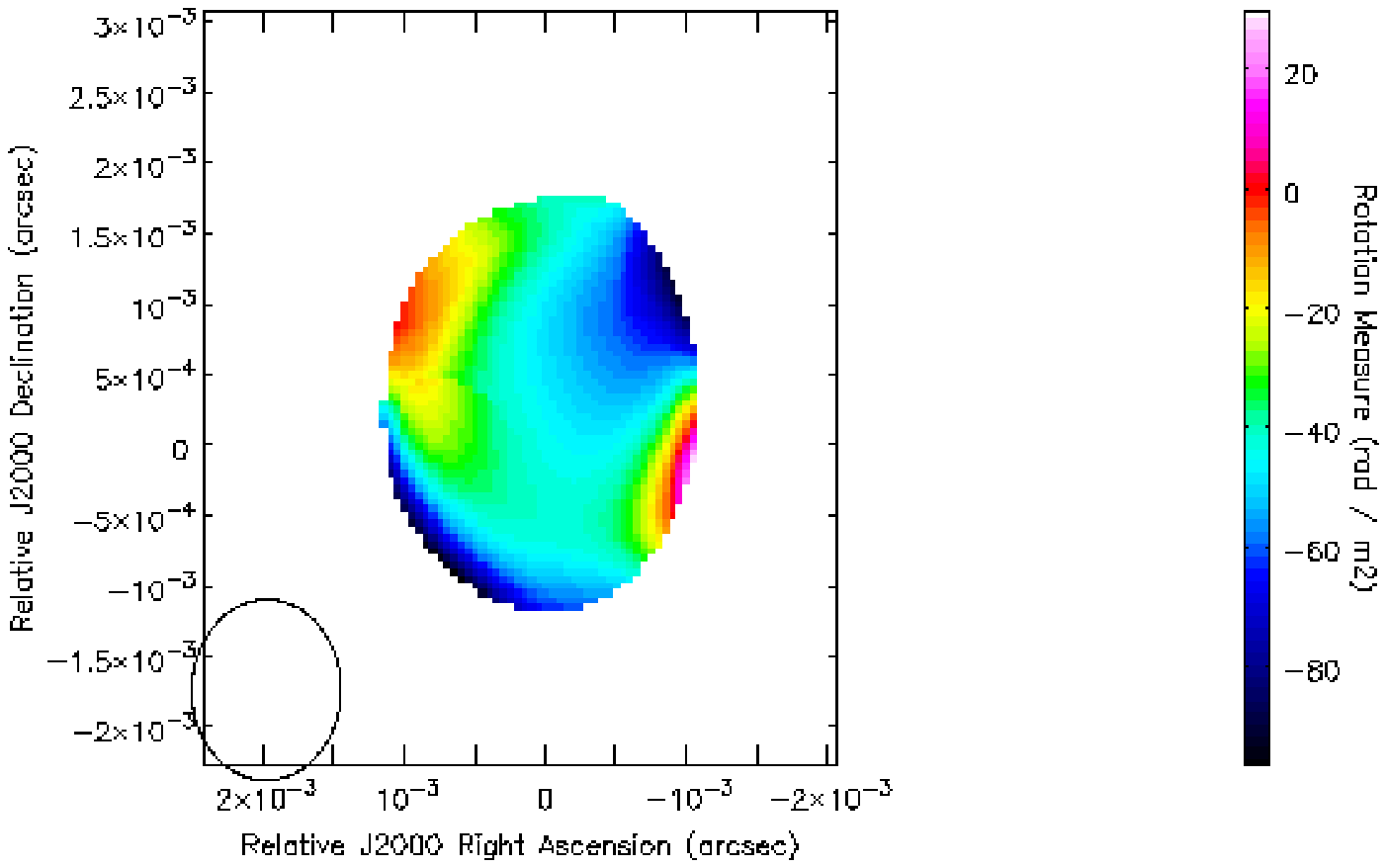}
 \end{center}
%\end{minipage}
\caption[Short caption for Figure 13]{\label{fig:MCG40} Same as Fig.~12
for a core--jet source with the same length but an intrinsic jet width
of 0.20~mas.  }
\end{figure}

\begin{figure}
% \begin{minipage}[t]{8.7cm}
 \begin{center}
 \includegraphics[width=9.2cm,clip]{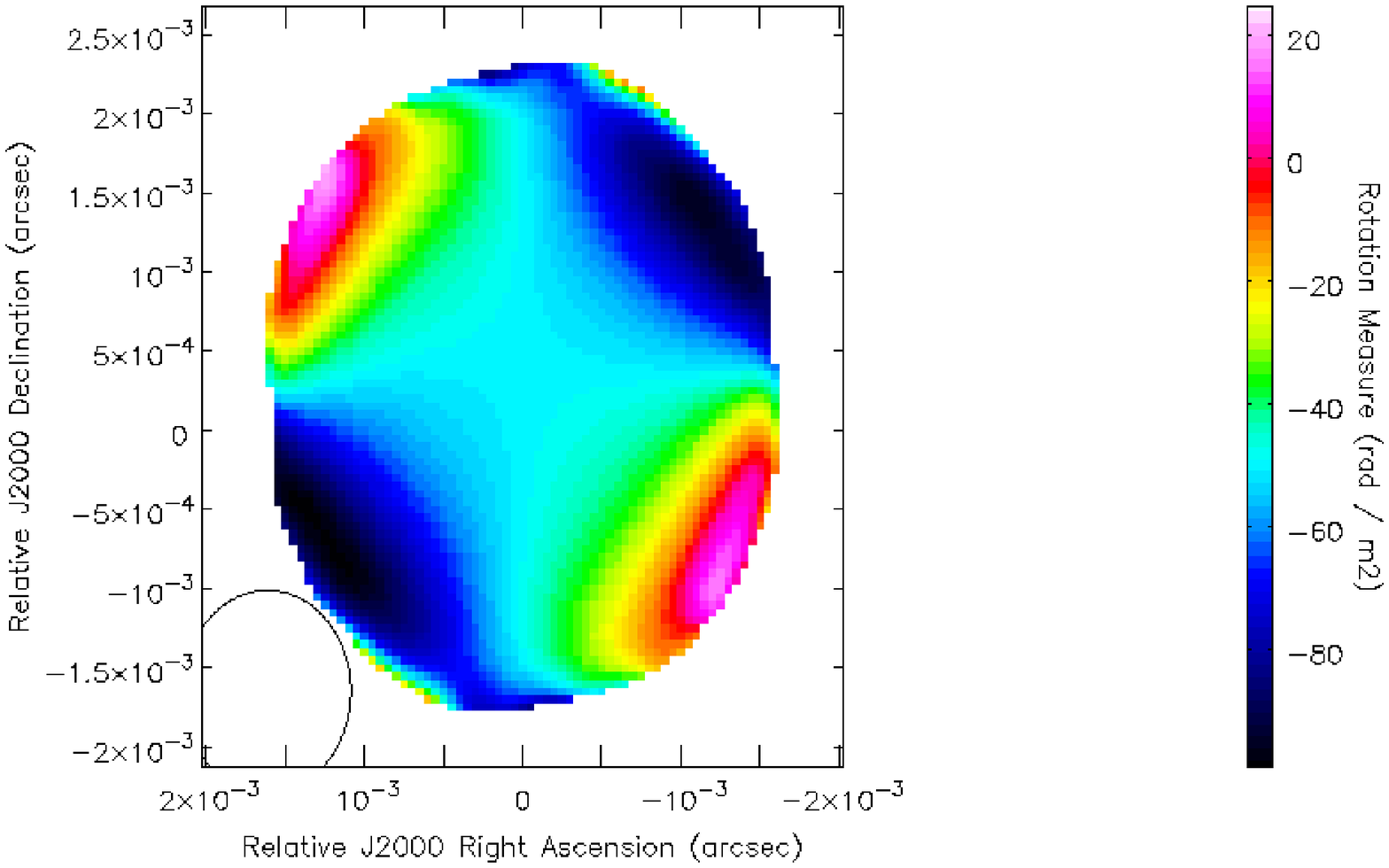}
 \end{center}
% \end{minipage}
% \begin{minipage}[t]{8.7cm}
% \begin{center}
% \includegraphics[width=8.7cm,clip]{MRCG20_RMAVG.eps}
% \end{center}
% \end{minipage}
%\begin{minipage}[t]{8.7cm}
 \begin{center}
 \includegraphics[width=9.2cm,clip]{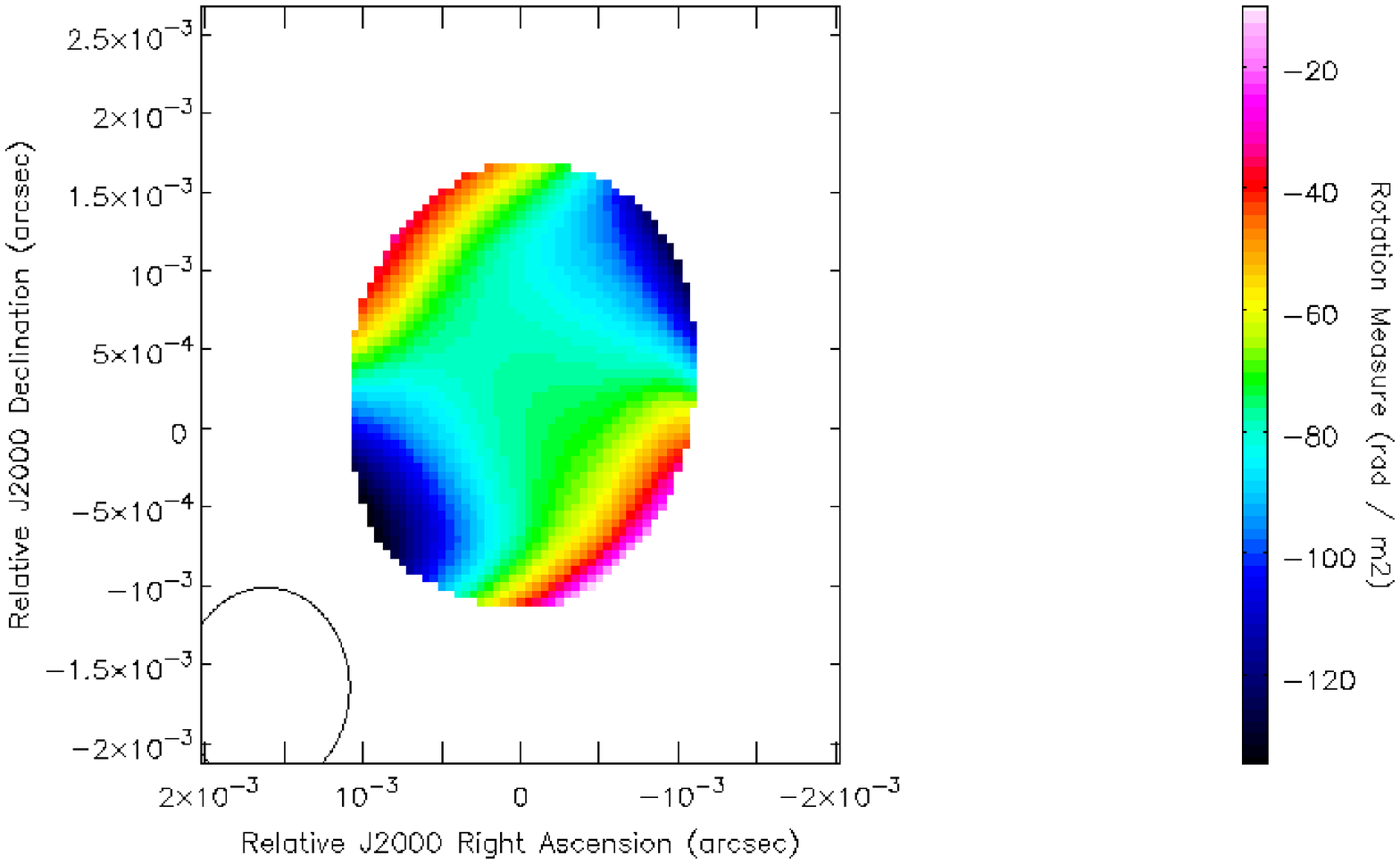}
 \end{center}
%\end{minipage}
%\begin{minipage}[t]{8.7cm}
 \begin{center}
 \includegraphics[width=9.2cm,clip]{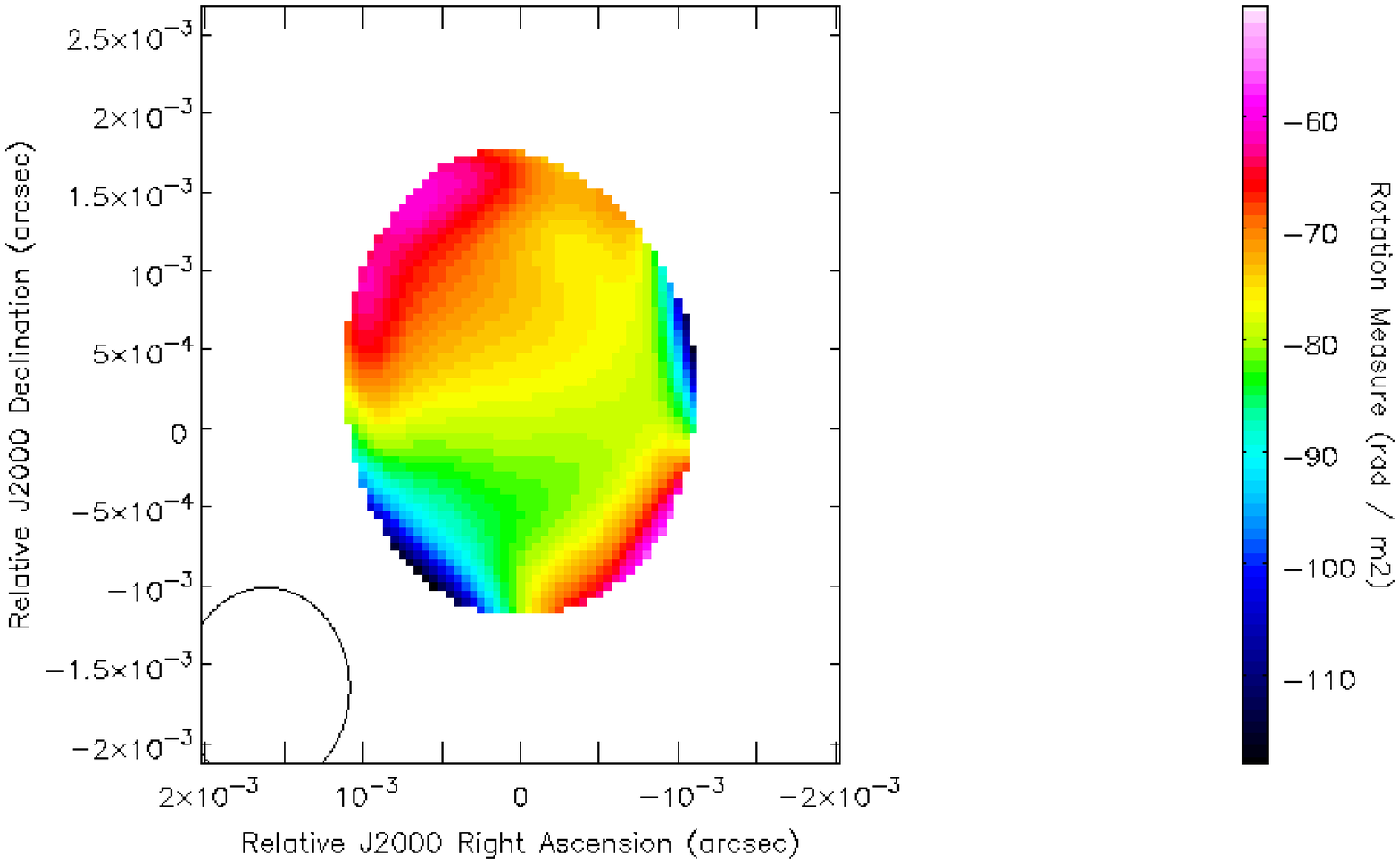}
 \end{center}
%\end{minipage}
\caption[Short caption for Figure 14]{\label{fig:MCG20} Same as Fig.~12
for a core--jet source with the same length but an intrinsic jet width
of 0.10~mas.  }
\end{figure}

\begin{figure}
% \begin{minipage}[t]{8.7cm}
 \begin{center}
 \includegraphics[width=9.2cm,clip]{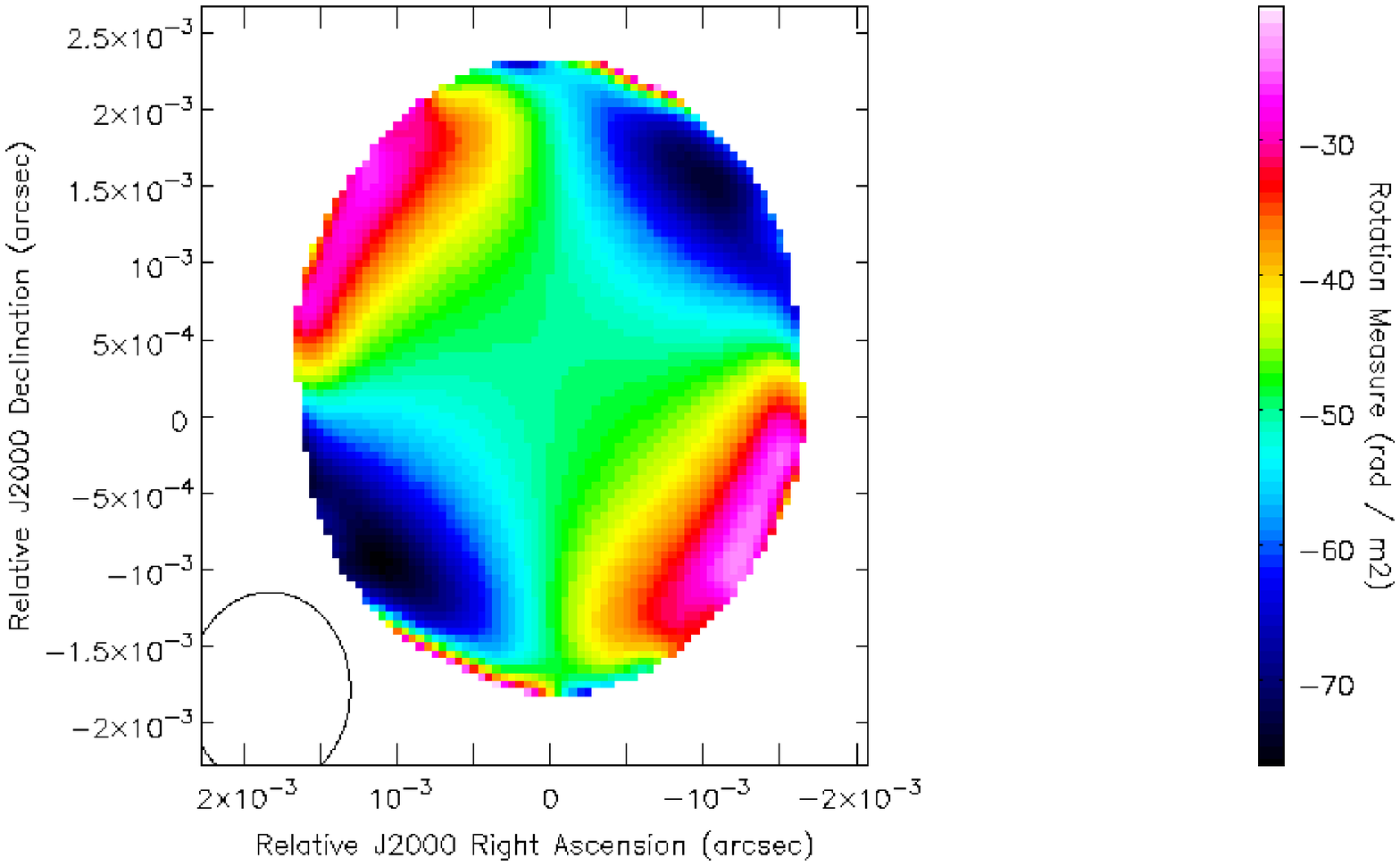}
 \end{center}
% \end{minipage}
% \begin{minipage}[t]{8.7cm}
% \begin{center}
% \includegraphics[width=8.7cm,clip]{MRCG10_RMAVG.eps}
% \end{center}
% \end{minipage}
%\begin{minipage}[t]{8.7cm}
 \begin{center}
 \includegraphics[width=9.2cm,clip]{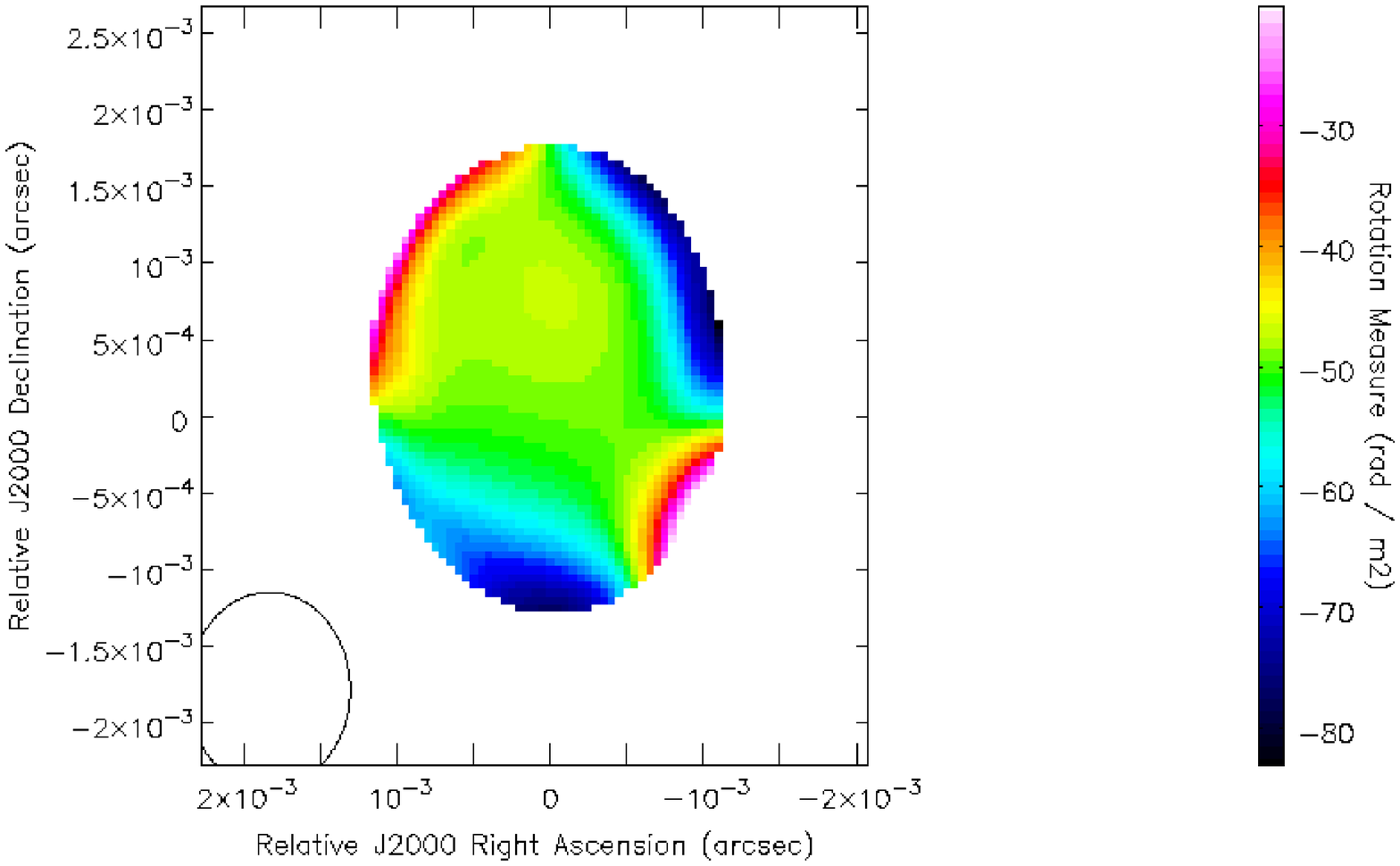}
 \end{center}
%\end{minipage}
%\begin{minipage}[t]{8.7cm}
 \begin{center}
 \includegraphics[width=9.2cm,clip]{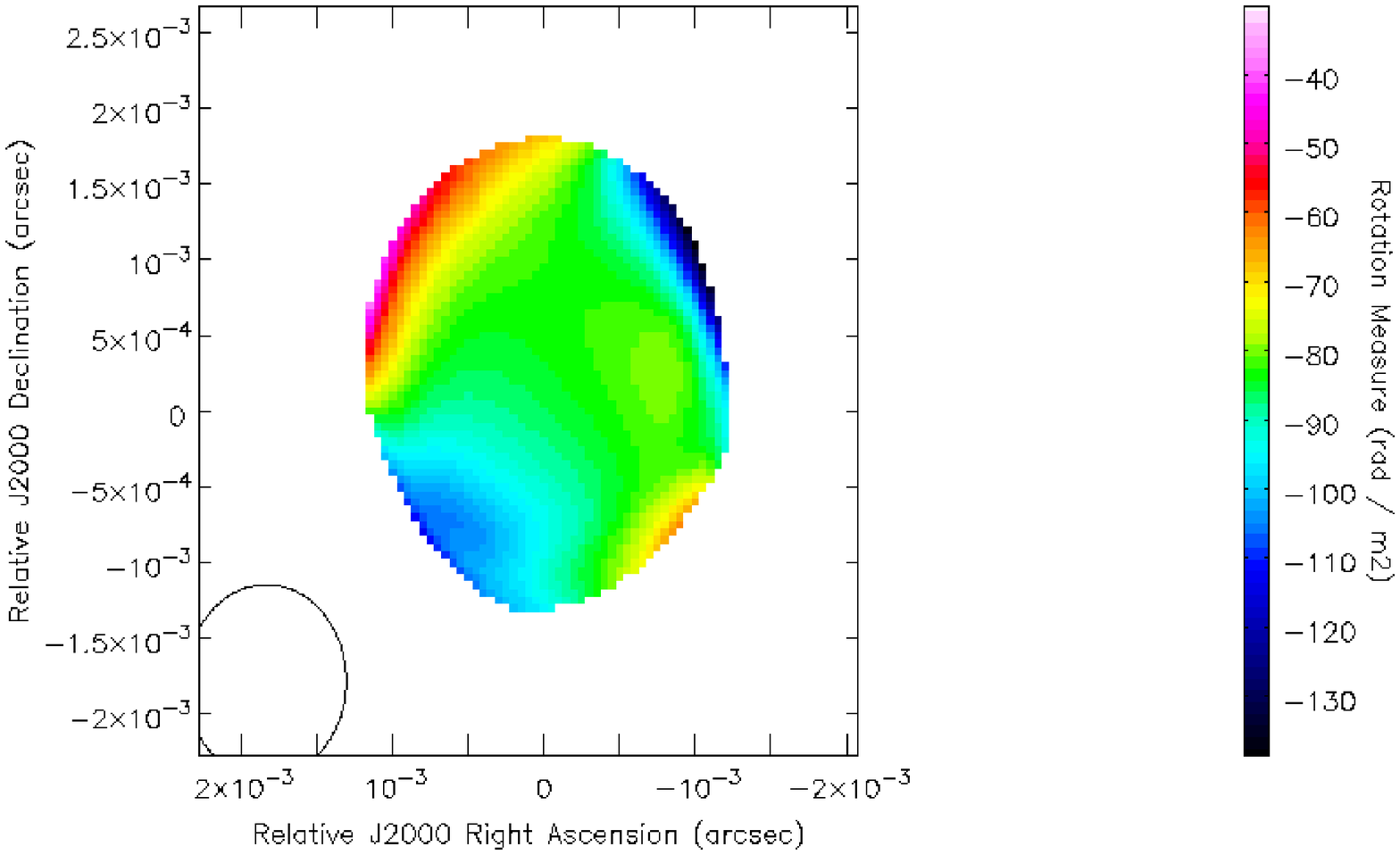}
 \end{center}
%\end{minipage}
\caption[Short caption for Figure 15]{\label{fig:MCG10} Same as Fig.~12
for a core--jet source with the same length but an intrinsic jet width
of 0.05~mas.  }
\end{figure}

\end{document}